\documentclass[12pt]{article}
\usepackage{enumerate}
\usepackage{amsfonts}
\usepackage{amsmath}
\usepackage{amssymb}
\usepackage{amsthm}
\usepackage{latexsym}
\usepackage{color}

\def\H{\mathcal{H}}

\def\S{\mathfrak{S}}
\def\C{\mathfrak{C}}
\def\T{\mathfrak{T}}
\def\B{\mathfrak{B}}

\newcommand{\id}{\mathrm{Id}}
\newcommand{\Tr}{\mathrm{Tr}}

\newcommand{\shs}{\hspace{1pt}}
\newcounter{defin}  \newcounter{lemma}  \newcounter{theorem}
\newcounter{property} \newcounter{corol}  \newcounter{remark} \newcounter{example}
\newenvironment{lemma}{\par\refstepcounter{lemma}     \textbf{Lemma \thelemma.} }{\rm\par}
\newenvironment{theorem}{\par\refstepcounter{theorem}     \textbf{Theorem \thetheorem.}\ }{\rm\par}
\newenvironment{property}{\par\refstepcounter{property}     \textbf{Proposition \theproperty.}\ }{\rm\par}
\newenvironment{corollary}{\par\refstepcounter{corol}     \textbf{Corollary \thecorol.} }{\rm\par}

\newenvironment{remark}{\par\refstepcounter{remark}     \textbf{Remark \theremark.}}{\rm\par}

\begin{document}

\title{Uniform continuity bounds for characteristics of multipartite quantum systems}
\author{M.E.~Shirokov\footnote{Steklov Mathematical Institute, Moscow, Russia, email:msh@mi.ras.ru}}
\date{}
\maketitle
\vspace{-30pt}
\begin{abstract}
We consider universal methods for obtaining (uniform) continuity bounds for characteristics of multipartite
quantum systems. We pay a special attention to infinite-dimensional multipartite
quantum systems under the energy constraints.

By these methods we obtain continuity bounds for several important characteristics of a multipartite quantum state: the quantum (conditional) mutual information, the squashed
entanglement, the c-squashed
entanglement and the conditional entanglement of
mutual information. The continuity bounds for the multipartite quantum  mutual information are asymptotically tight for large dimension/energy.

The obtained results are used to prove
the asymptotic continuity of the\break $n$-partite squashed entanglement,  c-squashed
entanglement and the conditional entanglement of
mutual information under the energy constraints.
\end{abstract}

\tableofcontents

\section{Introduction}

Multipartite quantum systems are basic objects in quantum information theory \cite{H-SCI,N&Ch,Wilde}. Such systems play central role in algorithms of quantum information processing, quantum computation, cryptography, etc.   Properties of states of multipartite quantum systems are described by different characteristics that are used essentially in analysis of  information abilities of such systems. So, important task consists in studying analytical properties of these characteristics
(as functions of a state), in particular, finding accurate upper and  lower estimates,  (uniform) continuity bounds (estimates for variation), conditions for asymptotic continuity (for entanglement measures), etc.

Continuity bounds for characteristics of a multipartite quantum state represented as a linear combination of the marginal entropies or conditional entropies of this state  can be obtained (in the finite-dimensional settings) by applying Audenaert's continuity bound for the entropy and Winter's continuity bound for the conditional entropy (cf. \cite{Aud,W-CB}) to each term of this linear combination.\footnote{Audenaert's continuity bound for the von Neumann entropy and Winter's continuity bound for the conditional entropy are optimized versions of the Fannes and Alicki-Fannes continuity bounds for these quantities \cite{A&F,Fannes}.} In the infinite-dimensional case the similar approach can be realized by means of Winter's continuity bounds for the entropy and the conditional entropy under the energy constraints \cite{W-CB}. The obvious drawback of this approach is low accuracy of the resulting continuity bounds. More accurate
continuity bounds for these characteristics can be obtained by  direct applications of the Alicki-Fannes-Winter method (\cite{A&F,W-CB}) and its infinite-dimensional generalizations (\cite{AFM,MCB})
to a characteristic of a multipartite quantum state without its decomposition.

 \smallskip

The aim of this paper is to consider universal methods  of obtaining accurate continuity bounds for characteristics of  multipartite quantum systems paying a special attention to infinite-dimensional
systems with the energy constraints of different forms.

 \smallskip

Mathematically, a characteristic of a multipartite quantum state is a function $f$ on the set $\S(\H_{A_1...A_n})$ of states of a
$n$-partite system  $A_1...A_n$, $n\geq 2$ (in infinite dimensions such function is typically well defined only on some subset of $\S(\H_{A_1...A_n})$). We will assume that
this function $f$ has the following property: $|f(\rho)|$ has an upper bound proportional to the sum of several marginal entropies of the state $\rho$. It means, w.l.o.g., that
\begin{equation}\label{F-p-2-p}
|f(\rho)|\leq C_f\sum_{k=1}^m H(\rho_{A_k}),\quad m\leq n,\quad C_f\in \mathbb{R}_+,
\end{equation}
for all states $\rho$ in $\S(\H_{A_1...A_n})$ having finite the marginal entropies
$H(\rho_{A_1})$,...,$H(\rho_{A_m})$ (for other states $\rho$ the function $f$ may not be defined). In fact, many real correlation
and entanglement measures on $\S(\H_{A_1...A_n})$ possess this property (see Sections 3,4).

In Section 3 we show that property (\ref{F-p-2-p}) is one of the conditions that allow to obtain continuity bound for the function $f$ valid for all states in $\S(\H_{A_1...A_n})$
with bounded energy corresponding to the system $A_1...A_m$. We note first that such continuity bound can be obtained by using the modification
of the Alicki-Fannes-Winter method proposed in \cite{AFM}, which is based on initial purification of quantum states with bounded energy. This approach gives simple and universal
continuity bounds for wide class of characteristics of  quantum systems composed of arbitrary subsystems provided that
\begin{equation}\label{H-cond-int}
 \lim_{\lambda\rightarrow0^+}\left[\mathrm{Tr}\, e^{-\lambda H_{A_k}}\right]^{\lambda}=1,\quad k=1,2,...,m,
\end{equation}
where $H_{A_k}$ is the Hamiltonian of the subsystem $A_k$ (Theorem \ref{main-gc}).\footnote{The sense of condition (\ref{H-cond-int}) is described in Section 2.2.} The main drawback of
continuity bounds obtained by this way is their non-accuracy for small distance between quantum  states.

More sharp universal continuity bounds can be obtained by using the two step technique based on appropriate finite-dimensional approximation
of states with bounded energy followed by the Alicki-Fannes-Winter method.\footnote{In the case $\,m=1\,$ this technique was used by A.Winter to obtain continuity
bounds for the entropy and the conditional entropy \cite{W-CB}.} The two step technique can be applied when the single subsystems $A_1$,...,$A_m$ are arbitrary
and their Hamiltonians satisfy condition (\ref{H-cond-int}),  but the
resulting continuity bounds are too complex in this case. So, to avoid technical difficulties and keeping in mind possible applications
we apply the two step technique assuming that the single subsystems $A_1$,...,$A_m$ (involved in (1))
are identical (it means that the
Hamiltonians $H_{A_1}$,...,$H_{A_m}$ of these subsystems are isomorphic). Under this assumption the
construction is simplified essentially (Theorem \ref{main}). We pay a special attention to the case when each of the subsystems $A_1$,...,$A_m$ is (isomorphic to)
a multi-mode quantum oscillator (Corolary \ref{SCB-G}).

In Section 4 we use general results of Section 3 to obtain continuity bounds for several important characteristics of a multipartite quantum state: the quantum (conditional) mutual information, the squashed
entanglement, the c-squashed entanglement and the conditional entanglement of mutual information. We show  that the continuity bounds for the multipartite quantum  mutual information  are asymptotically tight for large dimension/energy. We prove
the asymptotic continuity of the $n$-partite squashed entanglement, c-squashed
entanglement and the conditional entanglement of mutual information under the energy constraints.

In Section 5 we discuss an interesting feature of the proposed methods: the continuity bounds produced by these methods for many
characteristics of multipartite quantum systems remain valid after actions of any local channels on states of these systems.

\section{Preliminaries}

\subsection{Basic notations}

Let $\mathcal{H}$ be a separable Hilbert space,
$\mathfrak{B}(\mathcal{H})$ the algebra of all bounded operators on $\mathcal{H}$ with the operator norm $\|\cdot\|$ and $\mathfrak{T}( \mathcal{H})$ the
Banach space of all trace-class
operators on $\mathcal{H}$  with the trace norm $\|\!\cdot\!\|_1$. Let
$\mathfrak{S}(\mathcal{H})$ be  the set of quantum states (positive operators
in $\mathfrak{T}(\mathcal{H})$ with unit trace) \cite{H-SCI,N&Ch,Wilde}.

Denote by $I_{\mathcal{H}}$ the unit operator on a Hilbert space
$\mathcal{H}$ and by $\id_{\mathcal{\H}}$ the identity
transformation of the Banach space $\mathfrak{T}(\mathcal{H})$.\smallskip

The \emph{von Neumann entropy} of a quantum state
$\rho \in \mathfrak{S}(\H)$ is  defined by the formula
$H(\rho)=\operatorname{Tr}\eta(\rho)$, where  $\eta(x)=-x\ln x$ for $x>0$
and $\eta(0)=0$. It is a concave lower semicontinuous function on the set~$\mathfrak{S}(\H)$ taking values in~$[0,+\infty]$ \cite{H-SCI,L-2,W}.
The von Neumann entropy satisfies the inequality
\begin{equation}\label{w-k-ineq}
H(p\rho+(1-p)\sigma)\leq pH(\rho)+(1-p)H(\sigma)+h_2(p)
\end{equation}
valid for any states  $\rho$ and $\sigma$ in $\S(\H)$ and $p\in(0,1)$, where $\,h_2(p)=\eta(p)+\eta(1-p)\,$ is the binary entropy \cite{N&Ch,Wilde}.\smallskip

The \emph{quantum relative entropy} for two states $\rho$ and
$\sigma$ in $\mathfrak{S}(\mathcal{H})$ is defined as
$$
H(\rho\,\|\shs\sigma)=\sum\langle
i|\,\rho\ln\rho-\rho\ln\sigma\,|i\rangle,
$$
where $\{|i\rangle\}$ is the orthonormal basis of
eigenvectors of the state $\rho$ and it is assumed that
$H(\rho\,\|\sigma)=+\infty$ if $\,\mathrm{supp}\rho\shs$ is not
contained in $\shs\mathrm{supp}\shs\sigma$ \cite{H-SCI,L-2}.\footnote{The support $\mathrm{supp}\rho$ of a state $\rho$ is the closed subspace spanned by the eigenvectors of $\rho$ corresponding to its positive eigenvalues.}\smallskip

The \emph{quantum conditional entropy}
\begin{equation*}
H(A|B)_{\rho}=H(\rho)-H(\rho_{\shs B})
\end{equation*}
of a  state $\rho$ of a
bipartite quantum system $AB$ with finite marginal entropies is essentially used in analysis of quantum systems \cite{H-SCI,Wilde}. It
can be extended to the set of all states $\rho$ with finite $H(\rho_A)$ by the formula
\begin{equation*}
H(A|B)_{\rho}=H(\rho_{A})-H(\rho\shs\Vert\shs\rho_{A}\otimes
\rho_{B})
\end{equation*}
proposed in \cite{Kuz}.  This extension  possesses all basic properties of the quantum conditional entropy valid in finite dimensions \cite{Kuz,CMI}.  \smallskip

The \emph{quantum mutual information} of a state $\,\rho\,$ of a
bipartite quantum system $AB$ is defined as
\begin{equation}\label{mi-d}
I(A\!:\!B)_{\rho}=H(\rho\shs\Vert\shs\rho_{A}\otimes
\rho_{\shs B})=H(\rho_{A})+H(\rho_{\shs B})-H(\rho),
\end{equation}
where the second formula is valid if $\,H(\rho)\,$ is finite \cite{L-mi}.

The \emph{quantum conditional mutual information (QCMI)} of a state $\rho$ of a
tripartite finite-dimensional system $ABC$ is defined as
\begin{equation}\label{cmi-d}
    I(A\!:\!B|C)_{\rho}\doteq
    H(\rho_{AC})+H(\rho_{\shs BC})-H(\rho)-H(\rho_{C}).
\end{equation}
This quantity plays important role in quantum
information theory \cite{D&J,Wilde}, its nonnegativity is a basic result well known as \emph{strong subadditivity
of von Neumann entropy} \cite{Ruskai}. If system $C$ is trivial then (\ref{cmi-d}) coincides with (\ref{mi-d}).\smallskip

In infinite dimensions formula (\ref{cmi-d}) may contain the uncertainty
$"\infty-\infty"$. Nevertheless the
conditional mutual information can be defined for any state
$\rho$ in $\S(\H_{ABC})$ by the expression
\begin{equation}\label{cmi-e+}
I(A\!:\!B|C)_{\rho}=\sup_{P_A}\left[\shs I(A\!:\!BC)_{Q_A\rho
Q_A}-I(A\!:\!C)_{Q_A\rho Q_A}\shs\right],\;\; Q_A=P_A\otimes I_{BC},\!
\end{equation}
where the supremum is over all finite rank projectors
$P_A\in\B(\H_A)$ and it is assumed that $I(A\!:\!B')_{Q_A\rho
Q_A}=\lambda I(A\!:\!B')_{\lambda^{-1} Q_A\rho
Q_A}$, where $\lambda=\Tr\shs Q_A\rho$ \cite{CMI}.\smallskip

Expression (\ref{cmi-e+}) defines the  lower semicontinuous nonnegative function on the set
$\S(\H_{ABC})$ coinciding with the r.h.s. of (\ref{cmi-d}) for any state $\rho$ at which it is well defined and  possessing all basic properties of the quantum conditional mutual information valid in finite dimensions \cite[Th.2]{CMI}. In particular,
\begin{equation}\label{CMI-UB}
I(A\!:\!B|C)_{\rho}\leq 2\min\left\{H(\rho_A),H(\rho_{\shs B}),H(\rho_{AC}),H(\rho_{\shs BC}) \right\}
\end{equation}
for arbitrary state $\rho$ in $\S(\H_{ABC})$.\smallskip

The QCMI of a state $\,\rho\,$ of a finite-dimensional multipartite system $A_1 \ldots
A_nC$ is defined as follows (cf.\cite{AHS,Herbut,NQD,Y&C,YHW})
\begin{equation}\label{cmi-mpd}
\begin{array}{cl}
     I(A_1\!:\ldots:\!A_n|C)_{\rho}\!\!&\doteq\displaystyle
    \sum_{k=1}^n H(A_k|C)_{\rho}-H(A_1 \ldots A_n|C)_{\rho}\\&
    =\displaystyle\sum_{k=1}^{n-1} H(A_k|C)_{\rho}-H(A_1 \ldots A_{n-1}|A_nC)_{\rho}.
\end{array}
\end{equation}
Its nonnegativity and other basic properties can be derived from the
corresponding properties of the tripartite QCMI by using the representation (cf.\cite{Y&C})
\begin{equation}\label{cmi-mpd+}
\begin{array}{rl}
     I(A_1\!:\ldots:\!A_n|C)_{\rho}=I(A_{n-1}\!:\!A_n|C)_{\rho}\!\!& +\; I(A_{n-2}\!:\!A_{n-1}A_n|C)_{\rho}+...\\\\&+\;
     I(A_1\!:\!A_2...A_{n}|C)_{\rho}.
\end{array}
\end{equation}

By using representation (\ref{cmi-mpd+}) and the extended tripartite
QCMI described before one can define QCMI for any state of
an infinite-dimensional system $A_1...A_nC$.
The extended QCMI is a lower semicontinuous nonnegative function on the set
$\,\S(\H_{A_1 \ldots A_nC})$  coinciding with the r.h.s. of (\ref{cmi-mpd}) for any state $\rho$ in $\S(\H_{A_1...A_nC})$ with finite marginal entropies
and possessing basic properties of QCMI \cite[Proposition 5]{CMI}.\smallskip

If $\rho$ and $\sigma$ are states in $\S(\H_{A_1...A_nC})$ such that  $R=I(A_1\!:\ldots:\!A_n|C)_{\rho}$ and $S=I(A_1\!:\ldots:\!A_n|C)_{\sigma}$ are finite then
\begin{equation}\label{F-c-b+}
-h_2(p)\leq I(A_1\!:\ldots:\!A_n|C)_{p\rho+(1-p)\sigma}-[pR+(1-p)S]\leq (n-1) h_2(p)
\end{equation}
for any $p\in(0,1)$, where $h_2(p)$ is the binary entropy. Indeed, if $\rho$ and $\sigma$ are states with finite marginal
entropies then inequality (\ref{F-c-b+})
can be proved by using the second expression in (\ref{cmi-mpd}), concavity of the conditional entropy and  inequality (\ref{w-k-ineq}).
The validity of (\ref{F-c-b+}) for arbitrary states $\rho$ and $\sigma$ with finite QCMI can be shown by
approximation using Proposition 5 in \cite{CMI}.\smallskip

\subsection{The set of quantum states with bounded energy}\label{sec:22}

Let $H_A$ be a positive (semi-definite) densely defined operator on a Hilbert space $\mathcal{H}_A$.  We will assume that
$\Tr H_A\rho=\sup_n\mathrm{Tr} P_n H_A\rho$ for any positive operator $\rho\in\T(\H_A)$, where $P_n$ is the spectral projector of $H_A$ corresponding to the interval $[0,n]$. \smallskip

Let $E^A_0$ be the infimum of the spectrum of $H_A$ and $E\geq E^A_0$. Then
$$
\mathfrak{C}_{H_A,E}=\left\{\rho\in\mathfrak{S}(\mathcal{H}_A)\,|\,\mathrm{Tr} H_A\rho\leq E\right\}
$$
is a closed convex subset of $\mathfrak{S}(\mathcal{H}_A)$. If
$H_A$ is treated as  Hamiltonian of a quantum system $A$ then
$\mathfrak{C}_{H_A,E}$  is the set of states with the mean energy not exceeding $E$.\smallskip

It is well known that the von Neumann entropy is continuous on the set $\mathfrak{C}_{H_A,E}$ for any $E> E^A_0$ if (and only if) the Hamiltonian  $H_A$ satisfies  the condition
\begin{equation}\label{H-cond}
  \mathrm{Tr}\, e^{-\lambda H_{A}}<+\infty\quad\textrm{for all}\;\lambda>0
\end{equation}
and that the maximal value of the entropy on this set is achieved at the \emph{Gibbs state} $\gamma_A(E)\doteq e^{-\lambda(E) H_A}/\mathrm{Tr} e^{-\lambda(E) H_A}$, where the parameter $\lambda(E)$ is determined by the equality $\mathrm{Tr} H_A e^{-\lambda(E) H_A}=E\mathrm{Tr} e^{-\lambda(E) H_A}$ \cite{W}. Condition (\ref{H-cond}) implies that $H_A$ is an unbounded operator having  discrete spectrum of finite multiplicity. So, by the Lemma in \cite{H-c-w-c} the set $\mathfrak{C}_{H_A,E}$ is compact for any $E> E^A_0$.\footnote{The compactness of $\mathfrak{C}_{H_A,E}$ also follows from Corollary 7 in \cite{EC} which states that boundedness of the entropy on a convex set of quantum states implies  relative compactness of this set.}

We will use the function
\begin{equation}\label{F-def}
F_{H_A}(E)\doteq\sup_{\rho\in\mathfrak{C}_{H_{\!A},E}}H(\rho)=H(\gamma_A(E)).
\end{equation}
It is easy to show that $F_{H_A}$ is a strictly increasing concave function on $[E^A_0,+\infty)$ such that $F_{H_A}(E^A_0)=\ln m(E^A_0)$, where $m(E^A_0)$ is the multiplicity of $E^A_0$ \cite{EC,W-CB}.

In this paper we will assume that the Hamiltonian $H_A$ satisfies the condition
\begin{equation}\label{H-cond+}
  \lim_{\lambda\rightarrow0^+}\left[\mathrm{Tr}\, e^{-\lambda H_A}\right]^{\lambda}=1,
\end{equation}
which is slightly stronger than condition (\ref{H-cond}).\footnote{In terms of the sequence $\{E_k\}$ of eigenvalues of $H_A$
condition (\ref{H-cond}) means that $\lim_{k\rightarrow\infty}E_k/\ln k=+\infty$, while condition (\ref{H-cond+}) is valid  if $\;\liminf_{k\rightarrow\infty} E_k/\ln^q k>0\,$ for some $\,q>2$ \cite[Proposition 1]{AFM}.}
By Lemma 1 in \cite{AFM} condition (\ref{H-cond+}) holds if and only if
\begin{equation}\label{H-cond++}
  F_{H_A}(E)=o\shs(\sqrt{E})\quad\textrm{as}\quad E\rightarrow+\infty,
\end{equation}
while condition (\ref{H-cond}) is equivalent to  $\,F_{H_A}(E)=o\shs(E)\,$ as $\,E\rightarrow+\infty$ \cite{EC}.
It is essential that condition (\ref{H-cond+})  holds for the Hamiltonians of many real quantum systems \cite{Datta,AFM}.\footnote{Theorem 3 in \cite{Datta} shows that $\,F_{H_A}(E)=O\shs(\ln E)$ as $E\rightarrow+\infty\,$ if condition (\ref{BD-cond}) below holds.}

The function
\begin{equation}\label{F-bar}
 \bar{F}_{H_A}(E)=F_{H_A}(E+E^A_0)=H(\gamma_A(E+E^A_0))
\end{equation}
is concave and nondecreasing on $[0,+\infty)$. Let $\hat{F}_{H_A}$ be a continuous function
on $[0,+\infty)$ such that
\begin{equation}\label{F-cond-1}
\hat{F}_{H_A}(E)\geq \bar{F}_{H_A}(E)\quad \forall E>0,\quad \hat{F}_{H_A}(E)=o\shs(\sqrt{E})\quad\textrm{as}\quad E\rightarrow+\infty
\end{equation}
and
\begin{equation}\label{F-cond-2}
\hat{F}_{H_A}(E_1)<\hat{F}_{H_A}(E_2),\quad\hat{F}_{H_A}(E_1)/\sqrt{E_1}\geq \hat{F}_{H_A}(E_2)/\sqrt{E_2}\qquad \forall E_2>E_1>0.
\end{equation}
Sometimes we will additionally assume that
\begin{equation}\label{F-cond-3}
\hat{F}_{H_A}(E)=\bar{F}_{H_A}(E)(1+o(1))\quad\textrm{as}\quad E\to+\infty.
\end{equation}
The existence of a function $\hat{F}_{H_A}$ with the required properties is established in the following proposition proved in \cite{MCB}.\smallskip

\begin{property}\label{add-l}
A) \emph{If the Hamiltonian $H_A$ satisfies condition (\ref{H-cond+}) then
\begin{equation*}
\hat{F}^{*}_{H_A}(E)\doteq \sqrt{E}\sup_{E'\geq E}\bar{F}_{H_A}(E')/\sqrt{E'}
\end{equation*}
is the minimal function satisfying all the conditions in (\ref{F-cond-1}) and (\ref{F-cond-2}).}

B) \emph{Let
\begin{equation*}
\!N_{\shs\uparrow}[H_A](E)\doteq \sum_{k,j: E_k+E_j\leq E} E_k^2\quad \textrm{and}\quad N_{\downarrow}[H_A](E)\doteq \sum_{k,j: E_k+E_j\leq E} E_kE_j
\end{equation*}
for any $E>E^A_0$. If
\begin{equation}\label{BD-cond}
\exists \lim_{E\rightarrow+\infty}N_{\shs\uparrow}[H_A](E)/N_{\downarrow}[H_A](E)=a>1
\end{equation}
then
\begin{itemize}
  \item  there is $E_*$ such that the function $E\mapsto\bar{F}_{H_A}(E)/\sqrt{E}$ is nonincreasing for all $E\geq E_*$ and hence $\hat{F}^{*}_{H_A}(E)=\bar{F}_{H_A}(E)$ for all $E\geq E_*$;
  \item $\hat{F}^{*}_{H_A}(E)=(a-1)^{-1} (\ln E)(1+o(1))$ as $E\rightarrow+\infty$.
\end{itemize}}
\end{property}

Condition (\ref{BD-cond}) is valid for the Hamiltonians
of many real quantum systems \cite{Datta}.\smallskip

Practically, it is convenient to use functions $\hat{F}_{H_A}$ defined by simple formulae. The example of
such function $\hat{F}_{H_A}$ satisfying all the conditions in (\ref{F-cond-1}),(\ref{F-cond-2}) and (\ref{F-cond-3}) in the case when $A$ is a multimode quantum oscillator is considered in Section 3.2.\smallskip

We will use the following simple\smallskip

\begin{lemma}\label{u-est}\emph{ Let $\,H$ be a positive operator on a Hilbert space $\H$ having discrete spectrum of finite multiplicity  and $P_d$ the projector on the subspace $\H_d$ corresponding to the minimal $\,d$ eigenvalues $\,E_0,..,E_{d-1}$ of $H$ (taking the multiplicity into account). Then for any state $\rho\in\S(\H)$ such that $\Tr H\rho\leq E$ the following inequality holds}
$$
\Tr(I_{\H}-P_d)\rho\leq (E-E_0)/(E_d-E_0).
$$
\end{lemma}\smallskip

\emph{Proof.} Since $\Tr(I_{\H}-P_d)\rho=1-\Tr P_d\rho$, the required inequality follows directly from the inequalities 
$E_0\Tr P_d\rho\leq\Tr P_d H\rho$ and $E_d\Tr(I_{\H}-P_d)\rho\leq\Tr(I_{\H}-P_d)H\rho$. $\square$

\section{The main results}

\subsection{The finite-dimensional case}

Many important characteristics of states of a $n$-partite
finite-dimensional quantum system $A^n\doteq A_{1}...A_{n}$ have a form of a function $f$ on the set $\S(\H_{A^n})$
satisfying inequality (\ref{F-p-2-p}) for some $m\leq n$ and the inequalities
\begin{equation}\label{F-p-1}
-a_f h_2(p)\leq f(p\rho+(1-p)\sigma)-p f(\rho)-(1-p)f(\sigma)\leq b_fh_2(p)
\end{equation}
for any states $\rho$  and $\sigma$ in $\S(\H_{A^n})$ and any $p\in[0,1]$, where $h_2$ is the binary entropy (defined after (\ref{w-k-ineq})) and $a_f,b_f\in \mathbb{R}_+$. Inequality (\ref{F-p-2-p}) can be written in the following more accurate form:
\begin{equation}\label{F-p-2}
-c^-_f S_m(\rho)\leq f(\rho)\leq c^+_f S_m(\rho),\quad \textrm{where}\quad S_m(\rho)=\sum_{k=1}^m H(\rho_{A_k}),\;\; m\leq n,
\end{equation}
and $c^-_f,c^+_f\in \mathbb{R}_+$, for any state $\rho$ in $\S(\H_{A^n})$.

Let $L^n_m(C,D)$ be the class of functions on $\S(\H_{A^n})$
satisfying inequalities (\ref{F-p-1}) and (\ref{F-p-2}) with the parameters $a_f,b_f$ and $c^{\pm}_f$ such that $a_f+b_f=D$ and $c^-_f+c^+_f=C$ . Denote by $\widehat{L}^{m}_n(C,D)$ the class containing all functions in $L^{m}_n(C,D)$
and all functions of the form
$$
f(\rho)=\sup_{\lambda}f_{\lambda}(\rho)\quad \textrm{and} \quad f(\rho)=\inf_{\lambda}f_{\lambda}(\rho),
$$
where $\{f_{\lambda}\}$ is any family of functions in $L^{m}_n(C,D)$.

A noncomplete list of important entropic and information characteristics belonging to one of the classes $\widehat{L}^{m}_n(C,D)$ includes the von Neumann entropy, the conditional entropy, the $n$-partite quantum (conditional) mutual information, the one way classical correlation, the quantum discord, the mutual information of a quantum channel, the coherent information of a quantum channel, the information gain of a quantum measurement with and without quantum side information, the $n$-partite relative entropy of entanglement, the quantum topological entropy and its $n$-partite generalization. For example, the von Neumann entropy belongs to the class  $L^{1}_1(1,1)$, while the conditional entropy $H(A_1|A_2)$ lies in the class  $L^{1}_2(2,1)$. This can be shown easily by using concavity of the entropy and the conditional entropy, inequality (\ref{w-k-ineq}) and
the well known inequality $|H(A_1|A_2)_{\rho}|\leq H(\rho_{A_1})$. It's a little harder to show that the quantum discords $D(A_1|A_2)$ and $D(A_2|A_1)$ of a state of bipartite system $A_1A_2$
belong, respectively, to the classes $\widehat{L}^{1}_2(2,2)$ and $\widehat{L}^{1}_2(1,2)$ (we use the notation from \cite{Xi}).\smallskip

There is a general way to construct a characteristic $f$ of a $n$-partite
quantum system $A_{1}...A_{n}$ via appropriate characteristic $h$ of extended $(n+l)$-partite
quantum system $A_{1}...A_{n}A_{n+1}...A_{n+l}$: the value of $f$ at any state $\rho$ in $\S(\H_{A_{1}..A_{n}})$ is defined as
\begin{equation}\label{g-f}
f(\rho)\doteq \inf_{\hat{\rho}\in \mathfrak{M}(\rho)}h(\hat{\rho}),
\end{equation}
where $\mathfrak{M}(\rho)$ is a particular subset of the set
\begin{equation}\label{all-ext}
\{\hat{\rho}\in\S(\H_{A_{1}..A_{n+l}})\,|\,\hat{\rho}_{A_{1}..A_{n}}=\rho\}
\end{equation}
of all extensions  of the state $\rho$ to a state of $A_{1}...A_{n+l}$. For given $m\leq n$ we will denote by
$N^{m}_{n,s}(C,D)$ the class of all functions $f$ on $\S(\H_{A_{1}..A_{n}})$
defined by formula (\ref{g-f}) via particular function $h$ in $\widehat{L}^{m}_{n+l}(C,D)$ for some $l>0$ with $\mathfrak{M}(\rho)=\mathfrak{M}_s(\rho)$, $s=1,2,3$, where:
\begin{itemize}
  \item $\mathfrak{M}_1(\rho)$ is the set (\ref{all-ext}) of all extensions of $\rho$;
  \item $\mathfrak{M}_2(\rho)$ is the set of all extensions of $\rho$ having the form
  \begin{equation}\label{c-ext}
       \hat{\rho}=\sum_i p_i \rho_i\otimes |i\rangle\langle i|,
  \end{equation}
  where $\{\rho_i\}$ is a collection of states in $\S(\H_{A_{1}..A_{n}})$, $\{p_i\}$ is a probability distribution and $\{|i\rangle\}$ is an orthonormal basis in $\H_{A_{n+1}}$ (in this case $l=1$);
  \item $\mathfrak{M}_3(\rho)$ is the set of all extensions of $\rho$ having the form (\ref{c-ext}) in which $\{\rho_i\}$ is a collection of \emph{pure} states in $\S(\H_{A_{1}..A_{n}})$.
\end{itemize}

The classes $N^{m}_{n,s}(C,D)$ contain important entanglement measures obtained either by the convex roof construction or by the construction called "conditional entanglement" in \cite{CondEnt}. For example, the squashed entanglement, the c-squashed entanglement and the entanglement of formation in a bipartite system $A_1A_2$ belong, respectively, to the classes $N^{1}_{2,1}(1,1)$, $N^{1}_{2,2}(1,1)$ and $N^{1}_{2,3}(1,1)$.  Indeed, these characteristics can be expressed  by formula (\ref{g-f}) with $h(\hat{\rho})=\frac{1}{2}I(A_1\!:\!A_2|A_3)_{\hat{\rho}}$ in which the
infimum is taken, respectively, over the sets $\,\mathfrak{M}_1(\rho)$, $\mathfrak{M}_2(\rho)$ and $\mathfrak{M}_3(\rho)$.  It suffices to note that
the function $\varrho\mapsto\frac{1}{2}I(A_1\!:\!A_2|A_3)_{\varrho}$ belongs to the class  $L^1_3(1,1)$ by inequalities (\ref{CMI-UB}) and (\ref{F-c-b+}).

In $n$-partite  quantum systems, the squashed entanglement, the c-squashed entanglement, the conditional entanglement of
mutual information belong to one of the classes $N^{m}_{n,s}(C,D)$ (see details in Section 4). This also concerns other conditional entanglement measures obtained via some function from one of the classes $\widehat{L}^{m}_{n+l}(C,D)$, $l>0$ \cite{CondEnt,YHW}.\smallskip

The following proposition gives continuity bounds for functions from the classes  $\widehat{L}^{m}_{n}(C,D)$
and  $N^{m}_{n,s}(C,D)$ in the case of finite-dimensional subsystems $A_1$,...,$A_m$.\smallskip

\begin{property}\label{f-d-case} \emph{Let $f$
be a function on the set of states of a composite quantum system $A_1...A_m...A_n$, where the subsystems $A_1$,...,$A_m$ are finite-dimensional $(m\leq n)$.
Let $\rho$ and $\sigma$ be any states in $\S(\H_{A_1...A_n})$ such that $\;\frac{1}{2}\|\shs\rho-\sigma\|_1\leq\varepsilon\leq 1$.}

\emph{If the function $f$ belongs to the class $\widehat{L}^{m}_{n}(C,D)$ then
\begin{equation}\label{fcb}
 |f(\rho)-f(\sigma)|\leq C\varepsilon\ln \dim\H_{A_1...A_m}+Dg(\varepsilon),
\end{equation}
where $g(x)\doteq(1+x)h_2\!\left(\frac{x}{1+x}\right)=(x+1)\ln(x+1)-x\ln x$.}
\emph{If the function $f$ belongs to the class $N^{m}_{n,s}(C,D)$ then (\ref{fcb}) holds with
$\varepsilon$ replaced by $\sqrt{\varepsilon(2-\varepsilon)}$.}
\end{property}\smallskip

It will be shown in Section 4 that the continuity bounds given by Proposition \ref{f-d-case} for some important characteristics of multipartite quantum states are  asymptotically tight  (or close to tight) for large $\dim\H_{A_1...A_m}$ \cite{AT}. \smallskip

\emph{Proof.} If the function $f$ belongs to the class $L^{m}_{n}(C,D)$ then
it satisfies inequality (\ref{F-p-2}) with the parameters $c^-_f$ and $c^+_f$ such that $c^-_f+c^+_f=C$.
Since  the subsystems $A_1$,...,$A_m$ are finite-dimensional, it follows  that
$$
-c^-_f \dim\H_{A_1...A_m}\leq f(\rho)\leq c^+_f \dim\H_{A_1...A_m}
$$
 for any state $\rho$ in $\S(\H_{A_1...A_n})$. So, by applying the Alicki-Fannes-Winter method\footnote{The basic idea of this method is proposed in \cite{A&F}, it is then modified in \cite{M&H,SR&H,W-CB}.} (presented in the optimal form in \cite{W-CB} and
described in a full generality in the proof of Proposition 1 in \cite{CMI}) we obtain inequality (\ref{fcb}).
Since the r.h.s. of (\ref{fcb}) depends only on the parameters $C$,$D$ and $m$, this inequality remains valid for any function $f$ in $\widehat{L}^{m}_{n}(C,D)$.

Assume that $f$ is a  function from the class $N^{m}_{n,1}(C,D)$ defined via some function $h$ in $\widehat{L}^{m}_{n+l}(C,D)$. Then the standard arguments based on the isometrical
equivalence of all purifications of a given state (see \cite{C&W}) show that
\begin{equation}\label{g-f++}
f(\rho)=\inf_{\Lambda}\shs h(\id_{A_1...A_n}\otimes \Lambda(\bar{\rho})),
\end{equation}
where $\bar{\rho}$ is a given purification in $\,\S(\H_{A_1..A_nR})$ of the state $\rho$, i.e. a pure state such that $\Tr_R\shs\bar{\rho}=\rho$, and the infimum is over
all channels  $\Lambda:\T(\H_R)\rightarrow\T(\H_{A_{n+1}..A_{n+l}})$.

Since $\,\frac{1}{2}\|\shs\rho-\sigma\|_1\leq\varepsilon$, there exist purifications $\bar{\rho}$ and $\bar{\sigma}$ of the states $\rho$ and $\sigma$ such that
$\,\frac{1}{2}\|\shs\bar{\rho}-\bar{\sigma}\|_1\leq\delta\doteq\sqrt{\varepsilon(2-\varepsilon)}$ \cite{H-SCI,Wilde,W-CB}.  By monotonicity of the trace norm we have
\begin{equation}\label{norm-est}
\textstyle\frac{1}{2}\displaystyle\|\shs\id_{A_1...A_n}\otimes \Lambda(\bar{\rho})-\id_{A_1...A_n}\otimes \Lambda(\bar{\sigma})\|_1\leq\delta
\end{equation}
for any channel $\Lambda$. Thus, by applying continuity bound (\ref{fcb}) to the function $h$ we obtain
$$
|h(\id_{A_1...A_n}\otimes \Lambda(\bar{\rho}))-h(\id_{A_1...A_n}\otimes \Lambda(\bar{\sigma}))| \leq C\delta\ln \dim\H_{A_1...A_m}+Dg(\delta).
$$
Since the r.h.s. of this inequality does not depend on $\Lambda$, it follows from (\ref{g-f++}) that (\ref{fcb}) holds for the function $f$ with
$\varepsilon$ replaced by $\delta$.

Assume that $f$ is a  function from the class $N^{m}_{n,2}(C,D)$ defined via some function $h$ in $\widehat{L}^{m}_{n+l}(C,D)$. We may assume, w.l.o.g., that $f(\rho)\leq f(\sigma)$. For given $\epsilon>0$ let
$\hat{\rho}$ be an extension of $\rho$ having form (\ref{c-ext}) such that
\begin{equation}\label{t-ineq}
h(\hat{\rho})\leq f(\rho)+\epsilon.
\end{equation}
By the Schrodinger-Gisin–Hughston–Jozsa–Wootters theorem (cf.\cite{Schr,Gisin,HJW}) there is a pure state $\bar{\rho}$ in $\S(\H_{A_1...A_nR})$ and a POVM $\{M_i\}$
in $R$ such that $\rho=\Tr_R\shs\bar{\rho}$ and $p_i\rho_i=\Tr_R [I_{A_1...A_n}\otimes M_i]\shs\bar{\rho}\,$ for all $i$.  So, if
$$
\Lambda(\varrho)=\sum_i [\Tr M_i\varrho]|i\rangle\langle i|
$$
is a q-c channel from $R$ to $A_{n+1}$ then  $\hat{\rho}=\id_{A_1...A_n}\otimes \Lambda(\bar{\rho})$. Since $\,\frac{1}{2}\|\shs\rho-\sigma\|_1\leq\varepsilon$, there exists a pure state  $\bar{\sigma}$ in $\S(\H_{A_1...A_nR})$ such that $\sigma=\Tr_R\shs\bar{\sigma}$ and
$\,\frac{1}{2}\|\shs\bar{\rho}-\bar{\sigma}\|_1\leq\delta$ \cite{H-SCI,Wilde,W-CB}.
Since the inequality (\ref{norm-est}) holds for the above channel $\Lambda$  by monotonicity of the trace norm, by applying continuity bound (\ref{fcb}) to the function $h$ we obtain
$$
h(\hat{\rho})=h(\id_{A_1...A_n}\otimes \Lambda(\bar{\rho}))\geq h(\id_{A_1...A_n}\otimes \Lambda(\bar{\sigma}))-C\delta\ln \dim\H_{A_1...A_m}-Dg(\delta).
$$
Since the state $\,\id_{A_1...A_n}\otimes \Lambda(\bar{\sigma})\,$ belongs to the set $\mathfrak{M}_2(\sigma)$, this inequality and (\ref{t-ineq}) imply that
$$
f(\rho)\geq f(\sigma)-C\delta\ln \dim\H_{A_1...A_m}-Dg(\delta)-\epsilon.
$$

If $f$ is a  function from the class $N^{m}_{n,3}(C,D)$ then we can repeat the above arguments by noting that in this case the POVM $\{M_i\}$ consists of 1-rank operators, and hence
the state $\id_{A_1...A_n}\otimes \Lambda(\bar{\sigma})$ belongs to the set $\mathfrak{M}_3(\sigma)$. $\square$ \smallskip

Applications of Proposition \ref{f-d-case} to some important characteristics of multipartite
finite-dimensional quantum systems can be found in Section 4.

\subsection{The infinite-dimensional case: arbitrary subsystems}

Assume now that $A_1$,...,$A_n$ are arbitrary infinite-dimensional quantum systems. We denote by $L^{m}_n(C,D)$, $m\leq n$, the class of
all functions $f$ on the set
\begin{equation}\label{S-m}
\S_m(\H_{A_1..A_n})\doteq\left\{\rho\in\S(\H_{A_1..A_n})\,|\,H(\rho_{A_1}),..., H(\rho_{A_m})<+\infty\shs\right\}
\end{equation}
satisfying inequalities (\ref{F-p-1}) and (\ref{F-p-2}) with the nonnegative parameters
$a_f$, $b_f$ and $c^\pm_f$ such that $\,a_f+b_f=D\,$ and  $\,c^-_f+c^+_f=C$. The classes $\widehat{L}^{m}_n(C,D)$ and $N^{m}_{n,s}(C,D)$, $s=1,2,3$, are defined in the same way as in the finite-dimensional case (see the previous subsection).

We obtain continuity bounds for functions from the above classes under the energy constraint on the
system $A^m\doteq A_1...A_m$ assuming that the Hamiltonian of this system has the "standard" form
\begin{equation}\label{Hm}
H_{A^m}=H_{A_1}\otimes I_{A_2}\otimes...\otimes I_{A_m}+\cdots+I_{A_1}\otimes... \otimes I_{A_{m-1}}\otimes H_{A_m}.
\end{equation}
We will use the following simple observation.\smallskip

\begin{lemma}\label{sl} \emph{If $H_{\!A_1}$,.., $H_{\!A_m}$ are positive operators on the spaces $\H_{\!A_1}$,.., $\H_{\!A_m}$ satisfying condition (\ref{H-cond+}) then the operator $H_{A^m}$ on the space $\H_{A_1..A_m}$ defined in (\ref{Hm}) satisfies condition (\ref{H-cond+}) and }\footnote{Here and in what follows we use the notation introduced in Section 2.2.}
\begin{equation}\label{F-A-m}
  \bar{F}_{H_{\!A^m}}(E)\leq\bar{F}_{H_{\!A_1}}(E)+...+\bar{F}_{H_{\!A_m}}(E)\quad \forall E>0.
\end{equation}

\emph{If the operators $H_{\!A_1}$,.., $H_{\!A_m}$ are unitary equivalent to an operator $H_A$ on $\H_A$ then}
$$
\bar{F}_{H_{\!A^m}}(E)=m\bar{F}_{H_{\!A}}(E/m) \quad \forall E>0.
$$
\end{lemma}

\emph{Proof.} By the equivalence of (\ref{H-cond+}) and (\ref{H-cond++}) it suffices to prove inequality (\ref{F-A-m}).

By noting that $E_0^{A^m}=E_0^{A_1}+...+E_0^{A_m}$ we obtain
$$
\begin{array}{c}
\displaystyle\bar{F}_{H_{\!A^m}}(E)=F_{H_{\!A^m}}(E+E_0^{A^m})\leq\max_{E_1+...+E_m\leq E+E_0^{A^m}\!\!,\, E_i\geq E_0^{A_i}} [F_{H_{\!A_1}}(E_1)+...+F_{H_{\!A_m}}(E_m)]\\\\
\displaystyle=\max_{E_1+...+E_m\leq E,\, E_i\geq 0\,} [\bar{F}_{H_{\!A_1}}(E_1)+...+\bar{F}_{H_{\!A_m}}(E_m)]
\leq \bar{F}_{H_{\!A_1}}(E)+...+\bar{F}_{H_{\!A_m}}(E).
\end{array}
$$

If the operators $H_{\!A_1}$,.., $H_{\!A_m}$ are unitary equivalent to some operator $H_A$ then $\bar{F}_{H_{\!A_k}}(E)=\bar{F}_{H_{\!A}}(E)$, $k=\overline{1,m}$. So, the concavity
of the function $\bar{F}_{H_{\!A}}$ implies that the last maximum in the above inequality is attained at the point
$E_k=E/m$, $k=\overline{1,m}$. $\square$ \smallskip

The following theorem gives  continuity bounds for functions from the classes $\widehat{L}^{m}_n(C,D)$ and $N^{m}_{n,s}(C,D)$
under the energy constraint on the system $A^m\doteq A_1...A_m$.\smallskip

\begin{theorem}\label{main-gc} \emph{Let $H_{\!A_1}$,.., $H_{\!A_m}$ be positive operators on the spaces $\H_{\!A_1}$,.., $\H_{\!A_m}$ satisfying condition (\ref{H-cond+}) and $H_{A^m}$ the operator on the space $\H_{A_1..A_m}$ defined in (\ref{Hm}). Let
$\rho$ and $\sigma$ be arbitrary states in  $\S(\H_{A_1...A_n})$ such that $\,\sum_{k=1}^{m}\Tr H_{A_k}\rho_{A_k},\,\sum_{k=1}^{m}\Tr H_{A_k}\sigma_{A_k}\leq mE$ and $\;\frac{1}{2}\|\shs\rho-\sigma\|_1\leq\varepsilon\leq 1$. Then
\begin{equation}\label{SBC-ineq-gen}
    |f(\rho)-f(\sigma)|\leq C\sqrt{2\varepsilon}\bar{F}_{H_{A^m}}\!\!\left[\frac{m\bar{E}}{\varepsilon }\right]+Dg(\sqrt{2\varepsilon})
    \end{equation}
for any function $f$ from the class $\widehat{L}^{m}_n(C,D)$, where $\bar{E}=E-m^{-1}E_0^{A^m}$. Inequality (\ref{SBC-ineq-gen}) holds for any function $f$ from the class $N^{m}_{n,s}(C,D)$
with $\varepsilon$ replaced by $\sqrt{\varepsilon(2-\varepsilon)}$.}\footnote{The function $g(x)$ is defined after inequality (\ref{fcb}).}\smallskip

\emph{The right hand side of (\ref{SBC-ineq-gen}) tends to zero as $\,\varepsilon\to 0$.}
\end{theorem}\medskip

\begin{remark}\label{i-s} If the operators $H_{\!A_1}$,.., $H_{\!A_m}$ are unitary equivalent to some operator $H_A$ then the last assertion of Lemma \ref{sl} shows that inequality (\ref{SBC-ineq-gen}) can be rewritten as
\begin{equation}\label{SBC-ineq-gen+}
 |f(\rho)-f(\sigma)|\leq Cm\sqrt{2\varepsilon}\bar{F}_{H_{A}}\!\!\left[\frac{\bar{E}}{\varepsilon }\right]+Dg(\sqrt{2\varepsilon}).
\end{equation}
\end{remark}\smallskip

\begin{remark}\label{u-b} Replacing the function $\bar{F}_{H_{A^m}}$ by any its upper bound $\hat{F}_{H_{A^m}}$ such that the function
$E\mapsto \hat{F}_{H_{A^m}}(E)/\sqrt{E}$ is non-increasing makes inequality (\ref{SBC-ineq-gen}) valid for any $\varepsilon>0$ (including the case $\varepsilon>1$).
\end{remark}\smallskip

\emph{Proof of Theorem \ref{main-gc}.} Since $\Tr H_{A^m}[\rho_{A_1}\otimes...\otimes\rho_{A_m}]=\sum_{k=1}^m \Tr H_{A_k}\rho_{A_k}$, we have
$$
\sum_{k=1}^m H(\rho_{A_k})=H(\rho_{A_1}\otimes...\otimes\rho_{A_m})\leq F_{H_{\!A^m}}(mE)=\bar{F}_{H_{\!A^m}}(m\bar{E})
$$
for any state $\rho\in \S(\H_{A_1...A_n})$ such that $\Tr H_{A^m}\rho_{A^m}=\sum_{k=1}^m\Tr H_{A_k}\rho_{A_k}\leq mE$. Hence for any such state $\rho$  inequality (\ref{F-p-2}) implies that
\begin{equation}\label{F-p-2+}
-c_f^-\bar{F}_{H_{\!A^m}}(m\bar{E})\leq f(\rho)\leq c_f^+\bar{F}_{H_{\!A^m}}(m\bar{E}).
\end{equation}

Thus, in the case $\varepsilon<1/2$  inequality (\ref{SBC-ineq-gen}) for any function $f$ from the class $L^{m}_n(C,D)$ follows from Theorem 1 in \cite{AFM}. In the case $\varepsilon\geq 1/2$  this inequality directly follows from inequality (\ref{F-p-2+}). Since the r.h.s. of (\ref{SBC-ineq-gen}) depends only on the parameters $C$,$D$ and the characteristics of the operators $H_{\!A_1}$,..,$H_{\!A_m}$, this inequality remains valid for any function $f$ in $\widehat{L}^{m}_{n}(C,D)$.

If $f$ is a function from the class $N^{m}_{n,s}(C,D)$ then the validity of inequality (\ref{SBC-ineq-gen}) with $\varepsilon$ replaced by $\sqrt{\varepsilon(2-\varepsilon)}$ is proved by repeating the arguments used in the proof of Proposition  \ref{f-d-case}.\smallskip

The last assertion of the theorem follows from
Lemma \ref{sl}. $\square$
\smallskip


\begin{corollary}\label{main+cor} \emph{Let $H_{\!A_1}$,.., $H_{\!A_m}$ be positive operators on the spaces $\H_{\!A_1}$,.., $\H_{\!A_m}$ satisfying condition (\ref{H-cond+}). All
functions from the classes $\widehat{L}^{m}_n(C,D)$ and $N^{m}_{n,s}(C,D)$ are uniformly continuous on the set
\begin{equation*}
\left\{\rho\in\S(\H_{A_1...A_n})\,\left|\,\sum_{k=1}^{m}\Tr H_{A_k}\rho_{A_k}\leq E\right.\right\}
\end{equation*}
for any $E>E^{A^m}_0$, where $E^{A^m}_0$ is the sum of minimal eigenvalues of $H_{\!A_1}$,.., $H_{\!A_m}$.}
\end{corollary}\smallskip

\begin{remark}\label{main+s-r}  In the above analysis we considered the classes $\widehat{L}^{m}_{n}(C,D)$ of functions on the set
$\S_m(\H_{A_1..A_n})$ defined in (\ref{S-m}). In a similar way one can introduce the classes $\,\widehat{L}^{m}_{n}(C,D|\shs\S_0)\,$ of functions defined on arbitrary convex subset $\S_0$ of $\S_m(\H_{A_1..A_n})$. By Remark 2 in \cite{AFM} the above results (Theorem \ref{main-gc} and Corollary \ref{main+cor}) are generalized to functions from the
classes $\widehat{L}^{m}_{n}(C,D|\shs\S_0)$ provided that the set $\S_0$ has the following invariance property:
$$
\textrm{the states}\quad \frac{\Tr_R[\hat{\rho}-\hat{\sigma}]_{-}}{\Tr[\hat{\rho}-\hat{\sigma}]_{-}}\quad \textrm{and}\quad   \frac{\Tr_R[\hat{\rho}-\hat{\sigma}]_{+}}{\Tr[\hat{\rho}-\hat{\sigma}]_{+}}\quad \textrm{belong to the set}\;\S_0
$$
for arbitrary purifications $\hat{\rho}$ and $\hat{\sigma}$ in $\S(\H_{A_1..A_nR})$ of any different states $\rho$ and $\sigma$ in $\S_0$, where $T_{-}$ and $T_{+}$ are the negative and positive parts of a Hermitian operator $T$.

By the proof of Theorem 1 in \cite{AFM} the above invariance property  holds for any subset $\S_0$ of $\S_m(\H_{A_1..A_n})$ consisting of states $\rho$ with finite values
of given energy type functionals $\Tr H_1\rho$,..,$\Tr H_l\rho$, where $H_1$,..,$H_l$ are arbitrary positive operators on $\H_{A_1..A_n}$.
\end{remark}\smallskip

\subsection{The infinite-dimensional case: identical subsystems}

The continuity bounds given by Theorem \ref{main-gc} are simple and universal but non-accurate
for small $\varepsilon$ because of their dependance on $\sqrt{\varepsilon}$. More sharp universal
continuity bound can be obtained by using two step technique based on appropriate finite-dimensional approximation
of arbitrary states $\rho$ and $\sigma$ followed by the Alicki-Fannes-Winter method.\footnote{Similar technique was used by A.Winter in \cite{W-CB}.}

We apply the two step technique assuming that the subsystems $A_1$,...,$A_m$  (involved in (\ref{F-p-2})) are infinite-dimensional and isomorphic to a given system $A$. It means that the
Hamiltonians $H_{A_1}$,...,$H_{A_m}$ of these systems are unitary equivalent to the Hamiltonian $H_A$ of the system $A$. This assumption essentially simplifies the resulting continuity bound and seems reasonable from the point of view of potential applications.

In the following theorem we assume that the Hamiltonian $H_A$  satisfies condition (\ref{H-cond+}) and has minimal eigenvalue $E^A_0$.
We also assume that  $\hat{F}_{H_A}$ is any continuous function on $\mathbb{R}_+$ satisfying conditions (\ref{F-cond-1}) and  (\ref{F-cond-2}).\footnote{The role of $\hat{F}_{H_A}$ can  be played by the function $\hat{F}^*_{H_A}$ defined in Proposition \ref{add-l}.}
\smallskip
\begin{theorem}\label{main} \emph{Let $A^n\doteq A_1...A_n$, where $A_k\cong A$ for $k=\overline{1,m}$, $m\leq n$, and $A_{m+1},...,A_n$ are arbitrary systems. Let $\rho$ and $\sigma$ be arbitrary states in $\S(\H_{A^n})$ such that $\,\sum_{k=1}^{m}\Tr H_{A_k}\rho_{A_k},\,\sum_{k=1}^{m}\Tr H_{A_k}\sigma_{A_k}\leq mE$ and $\;\frac{1}{2}\|\shs\rho-\sigma\|_1\leq\varepsilon$. Let $\,t\in (0, 1/\varepsilon)$. Then
\begin{equation}\label{SBC-ineq}
\begin{array}{ccc}
\displaystyle|f(\rho)-f(\sigma)|\leq
Cm\!\left((\varepsilon+\varepsilon^2t^2)\hat{F}_{H_{\!A}}\!\!\left[\frac{m\bar{E}}{\varepsilon^2t^2}\right]+2\sqrt{2\varepsilon t}\hat{F}_{H_{\!A}}\!\!\left[\frac{\bar{E}}{\varepsilon t}\right]\right)\\\\
+D\!\left(g\!\left(\varepsilon+\varepsilon^2t^2\right)+2g(\sqrt{2\varepsilon t})\right)
\end{array}
\end{equation}
for any function $f$ from the class $\widehat{L}^{m}_n(C,D)$, where $\bar{E}=E-E^A_0$. Inequality (\ref{SBC-ineq}) holds for any function $f$ from the class $N^{m}_{n,s}(C,D)$
with $\varepsilon$ replaced by $\sqrt{\varepsilon(2-\varepsilon)}$.}\footnote{The function $g(x)$ is defined after inequality (\ref{fcb}).}\smallskip

\emph{If conditions (\ref{F-cond-3}) and (\ref{BD-cond}) hold~\footnote{By Proposition \ref{add-l} this holds, in particular, if  $\hat{F}_{H_A}=\hat{F}^*_{H_A}$.} then for given $\bar{E}$ the r.h.s. of  (\ref{SBC-ineq}) can be written as
\begin{equation}\label{SBC-ineq-a}
\begin{array}{ccc}
\displaystyle Cm\!\left((\varepsilon+\varepsilon^2t^2)\ln\!\left[\frac{m\bar{E}}{\varepsilon^2t^2}\right]\frac{1+o(1)}{a-1}+2\sqrt{2\varepsilon t}\ln\!\left[\frac{\bar{E}}{\varepsilon t}\right]\frac{1+o(1)}{a-1}\right)\\\\
\displaystyle+D\!\left(g\!\left(\varepsilon+\varepsilon^2t^2\right)+2g(\sqrt{2\varepsilon t})\right),\quad\varepsilon t\rightarrow0^+.
\end{array}
\end{equation}
If, in addition, $f$ is a function from the class $\widehat{L}^{m}_n(C,D)$ satisfying inequality (\ref{F-p-2}) with the parameters
$c^-_f$ and $c^+_f$ such that
\begin{equation}\label{at}
\!\lim_{E\rightarrow+\infty} \left[\frac{\inf_{\rho\in\C^m_{E}} f(\rho)}{mF_{H_A}(E)}+c^-_f\right]=\lim_{E\rightarrow+\infty}\left[c^+_f-\frac{\sup_{\rho\in\C^m_{E}}f(\rho)}{mF_{H_A}(E)}\right]=0,
\end{equation}
where $\C^m_{E}=\{\shs\rho\in\S(\H_{A^n})\shs|\,\sum_{k=1}^{m}\Tr H_{A_k}\rho_{A_k}\leq mE\shs\}$ and $F_{H_A}$ is the function defined in (\ref{F-def}), then continuity bound (\ref{SBC-ineq}) with optimal $\,t$ is  asymptotically tight for large $E$ \cite{AT}.}
\end{theorem}\smallskip

\begin{remark}\label{inc-p} Since the function $\hat{F}_{H_A}$ satisfies condition (\ref{F-cond-1}) and (\ref{F-cond-2}),  the r.h.s. of (\ref{SBC-ineq}) (denoted by $\mathbb{VB}^m_{\shs t}(\bar{E},\varepsilon\,|\,C,D)$ in what follows) is a nondecreasing function of $\varepsilon$ and $\bar{E}$ tending to zero as $\,\varepsilon\rightarrow0^+$ for each $m$ and any given $\bar{E}$, $C$, $D$ and $\,t\in(0,1/\varepsilon)$.
\end{remark}\smallskip

\begin{remark}\label{t-r}
 The "free" parameter $\,t\,$ can be used to optimize continuity bound (\ref{SBC-ineq}) for given values of $E$ and $\varepsilon$.
\end{remark}\smallskip

\emph{Proof.} Since the Hamiltonian $H_A$  satisfies condition (\ref{H-cond+}), it has discrete spectrum of finite multiplicity. So,
we may assume that
$$
H_{A_k}=\sum_{i=0}^{+\infty}E_i|\tau^k_i\rangle\langle \tau^k_i|,\quad k=\overline{1,m},
$$
where $\{\tau^k_i\}$ is an orthonormal basis in the Hilbert space $\H_{A_k}$ and
$\{E_i\}$ is a nondecreasing sequence of eigenvalues of $H_{A}$.
Let
$$
\bar{H}_{A_k}=H_{A_k}-E^A_0I_{A_k}=\sum_{i=0}^{+\infty}\bar{E}_i|\tau^k_i\rangle\langle \tau^k_i|,
$$
where $\bar{E}_i=E_i-E^A_0$, $P^k_d$
be the projector onto the  subspace of $\H_{A_k}$ spanned by the vectors $\tau^k_0,...\tau^k_{d-1}$ and
$\,\bar{P}^k_d=I_{A_k}-P^k_d\,$ the projector onto the orthogonal subspace.

For each $d$ such that $\bar{E}_d>m\bar{E}$  consider the states
$$
\rho_d=r^{-1}_d Q_d\rho\shs Q_d \quad \textrm{and} \quad \sigma_d=s^{-1}_d Q_d\sigma Q_d,
$$
where $Q_d=P_d^1\otimes...\otimes P_d^m\otimes I_{A_{m+1}}\otimes...\otimes I_{A_n}$,
\begin{equation}\label{p-ineq}
r_d\doteq\Tr Q_d\rho\geq 1-m\bar{E}/\bar{E}_d
\quad \textrm{and}\quad s_d\doteq\Tr Q_d\sigma\geq 1-m\bar{E}/\bar{E}_d.
\end{equation}
To prove the first inequality in (\ref{p-ineq}) note that Lemma \ref{u-est} in Section 2.2 implies
$$
\begin{array}{c}
\left|\Tr Q_d^{k-1}\rho - \Tr Q_d^{k}\rho\shs\right|
\leq \|Q_d^{k-1}\|\Tr[I_{A_1}\otimes...\otimes I_{A_{k-1}}\otimes \bar{P}_d^k\otimes I_{A_{k+1 }}\otimes...\otimes I_{A_{n}}]\rho\\\\
=\Tr\bar{P}_d^k\rho_{A_k}\leq \Tr \bar{H}_{A_k}\rho_{A_k}/\bar{E}_d,\quad k=\overline{1,m},
\end{array}
$$
where $Q_d^0=I_{A^n}$ and $\,Q_d^k=P_d^1\otimes...\otimes P_d^{k}\otimes I_{A_{k+1}}\otimes...\otimes I_{A_n}$, $k=\overline{1,m}$. It follows that
$$
1-r_d\leq \sum_{k=1}^m \left|\Tr Q_d^{k-1}\rho - \Tr Q_d^{k}\rho\shs\right|\leq \sum_{k=1}^m \Tr \bar{H}_{A_k}\rho_{A_k}/\bar{E}_d\leq m\bar{E}/\bar{E}_d.
$$
The second inequality in (\ref{p-ineq}) is proved similarly.\smallskip

The condition $\bar{E}_d>m\bar{E}$ implies that
\begin{equation}\label{e-cond}
  \sum_{k=1}^{m}\Tr H_{A_k}[\rho_d]_{A_k}\leq mE\quad\textrm{and}\quad \sum_{k=1}^{m}\Tr H_{A_k}[\sigma_d]_{A_k}\leq mE.
\end{equation}
Indeed, by the assumption we have
\begin{equation*}
  \Tr \bar{H}_{A^m}\rho_{A^m}\leq m\bar{E},
\end{equation*}
where $\bar{H}_{A^m}=H_{A^m}-mE^A_0 I_{A^m}$ ($H_{A^m}$ is the operator defined in (\ref{Hm})). Hence
$$
\Tr \bar{H}_{A^m}[\rho_d]_{A^m}\leq r_d^{-1}(m\bar{E}-\Tr \bar{H}_{A^m}T_d\rho_{A^m})\leq r_d^{-1}(m\bar{E}-\bar{E}_d\Tr T_d\rho_{A^m})\leq m\bar{E},
$$
where $T_d=I_{A^m}-P_d^1\otimes...\otimes P_d^{m}\,$  and the second inequality follows from the fact that all eigenvalues of  $\bar{H}_{A^m}$ corresponding to the
range of $T_d$ are not less than $\bar{E}_d$. The second inequality in (\ref{e-cond}) is proved similarly.

Winter's gentle measurement lemma (cf.\cite{W-gml,Wilde}) implies
\begin{equation}\label{t-norm-est}
\|\omega-\omega_d\|_1\leq 2\sqrt{\Tr \bar{Q}_d\omega}\leq2\sqrt{m\bar{E}/\bar{E}_d},\quad\omega=\rho,\sigma,
\end{equation}
where $\,\bar{Q}_d=I_{A^n}-Q_d\,$ and the last inequality follows from (\ref{p-ineq}).

Assume that $f$ is a function from the class $\widehat{L}^{m}_n(C,D)$. By using (\ref{e-cond}) and (\ref{t-norm-est}) we obtain from Theorem \ref{main-gc} with Remarks \ref{i-s} and \ref{u-b} that
\begin{equation}\label{d-one}
  |f(\rho)-f(\rho_d)|,|f(\sigma)-f(\sigma_d)|\leq Cm\sqrt{2\delta_d}\hat{F}_{H_A}(\bar{E}/\delta_d)+Dg(\sqrt{2\delta_d}),
\end{equation}
where $\delta_d=\sqrt{m\bar{E}/\bar{E}_d}$.

By using monotonicity of the trace norm under quantum operations and the inequalities in (\ref{p-ineq}) we obtain
\begin{equation*}
\begin{array}{c}
  \|\rho_d-\sigma_d\|_1\leq \|\shs Q_d\rho\shs Q_d-Q_d\sigma Q_d\|_1+\|Q_d\rho\shs Q_d\|_1|1- r_d^{-1}|+\|Q_d\sigma Q_d\|_1|1- s_d^{-1}|\\\\
  \leq 2\varepsilon+(1-r_d)+(1-s_d)\leq 2\varepsilon+2m\bar{E}/\bar{E}_d.
\end{array}
\end{equation*}
Thus, since the states $[\rho_d]_{A_k}$ and $[\sigma_d]_{A_k}$ are supported by the $d$-dimensional subspace  $P_d^k(\H_{A_k})$ for each $k=\overline{1,m}$,
it follows from Proposition  \ref{f-d-case} that
\begin{equation}\label{d-two}
  |f(\rho_d)-f(\sigma_d)|\leq Cm\varepsilon_d\ln d+Dg(\varepsilon_d),
\end{equation}
where $\varepsilon_d=\varepsilon+m\bar{E}/\bar{E}_d$.

By using inequalities (\ref{d-one}) and (\ref{d-two}) we obtain
\begin{equation}\label{p-CB}
\begin{array}{ccc}
|f(\rho)-f(\sigma)|\leq |f(\rho)-f(\rho_d)|+|f(\sigma)-f(\sigma_d)|+|f(\rho_d)-f(\sigma_d)|\\\\\leq
Cm\!\left(2\sqrt{2\delta_d}\hat{F}_{H_A}(\bar{E}/\delta_d)+\varepsilon_d\ln d\right)+D\!\left(2g(\sqrt{2\delta_d})+g(\varepsilon_d)\right).
\end{array}
\end{equation}
Since $\|H_{A_1}P^1_d\|=E_{d-1}$, we have
\begin{equation}\label{ln-d}
\ln d=H(d^{-1}P^1_d)\leq F_{H_{\!A_1}}(E_{d-1})=F_{H_{\!A}}(E_{d-1})=\bar{F}_{H_{\!A}}(\bar{E}_{d-1})\leq \hat{F}_{H_{\!A}}(\bar{E}_{d-1})\quad \forall d.
\end{equation}
If $m\bar{E}\geq \varepsilon^2 t^2\bar{E}_{d_0}$ for given $t\in(0,1/\varepsilon)$, where $d_0$ is the multiplicity of $E^A_0$, then there is $d_*>d_0$ such that $m\bar{E}<\bar{E}_{d_*}$ and
\begin{equation}\label{din}
 \frac{m\bar{E}}{\bar{E}_{d_*}}\leq \varepsilon^2t^2\leq \frac{m\bar{E}}{\bar{E}_{d_*-1}}.
\end{equation}
By using (\ref{ln-d}),  the second inequality in (\ref{din}) and the monotonicity of $\hat{F}_{H_A}$ we obtain
\begin{equation}\label{ln-d-}
\ln d_*\leq \hat{F}_{H_A}(m\bar{E}/(\varepsilon^2t^2)).
\end{equation}
If $m\bar{E}<\varepsilon^2 t^2\bar{E}_{d_0}$ then by setting $d_*=d_0$ we obtain the first inequality in (\ref{din}),
$$
m\bar{E}<\bar{E}_{d_*}\quad \textrm{and} \quad \ln d_*=F_{H_A}(E^A_0)=\bar{F}_{H_A}(0)\leq \hat{F}_{H_A}(0).
$$
So, by monotonicity of $\hat{F}_{H_A}$, inequality (\ref{ln-d-}) holds in this case as well.

By using the first inequality in (\ref{din}), upper bound (\ref{ln-d-}) and monotonicity of the
functions $E\mapsto \hat{F}_{H_A}(E)/\sqrt{E}$ and $g(x)$, it is easy to obtain inequality (\ref{SBC-ineq}) from the inequality (\ref{p-CB}) with $d=d_*$.
\smallskip

If $f$ is a function from the class $N^{m}_{n,s}(C,D)$ then the validity of inequality (\ref{SBC-ineq}) with $\varepsilon$ replaced by $\sqrt{\varepsilon(2-\varepsilon)}$ is proved by repeating the arguments used in the proof of Proposition  \ref{f-d-case}.\smallskip

Assume that conditions (\ref{F-cond-3}) and (\ref{BD-cond}) hold.  Then it follows from Proposition \ref{add-l}B
that
\begin{equation}\label{f-rel}
\hat{F}_{H_A}(E)=(a-1)^{-1}\ln(E)(1+o(1))\quad\textrm{as }\;E \to +\infty.
\end{equation}
This implies the asymptotic representation (\ref{SBC-ineq-a}).

Assume that $f$ is a function from the class $\widehat{L}^{m}_n(C,D)$ satisfying condition (\ref{at}). Then
for any $\delta>0$ there
exists $E_{\delta}>E_0^A$ such that for any $E>E_{\delta}$ the set $\C^m_{E}$ contains states $\rho$ and
$\sigma$ such that
$|f(\rho)-f(\sigma)|\geq (C-\delta)mF_{H_A}(E)$.
Since $\frac{1}{2}\|\rho-\sigma\|_1\leq 1$, it follows that for any $\varepsilon>0$ the set $\C^m_{E}$ contains states $\rho_{\varepsilon}$ and
$\sigma_{\varepsilon}$ such that\footnote{This can be shown by using the states $\rho_k=\frac{k}{n}\rho+(1-\frac{k}{n})\sigma$, $k=0,1,..,n$, for sufficiently large $n$.}
\begin{equation}\label{t-p}
\textstyle\frac{1}{2}\|\rho_{\varepsilon}-\sigma_{\varepsilon}\|_1\leq \varepsilon\quad \textrm{and}\quad |f(\rho_{\varepsilon})-f(\sigma_{\varepsilon})|\geq \varepsilon (C-\delta)mF_{H_A}(E).
\end{equation}

By using (\ref{f-rel}) and the similar representation for the function
$F_{H_A}(E)$ (Theorem 3 in \cite{Datta}) one can show that for any $\delta>0$ there exists $E_{\delta}>E_0^A$ and $\varepsilon_{\delta}\in(0,1)$ such that  the r.h.s. of (\ref{SBC-ineq}) with $t=\varepsilon^2$ does not exceed
\begin{equation}\label{rhs-r}
C\varepsilon m F_{H_A}(E)(1+\delta)+X(\varepsilon,E)\quad \textrm{for all}\quad E\geq E_{\delta}\;\;\textrm{and}\;\; \varepsilon\leq\varepsilon_{\delta},
\end{equation}
where $X(\varepsilon,E)$ is a bounded function.

Since $F_{H_A}(E)$ tends to $+\infty$ as $E\to+\infty$, by using upper bound (\ref{rhs-r}) and the  states $\rho_{\varepsilon}$ and
$\sigma_{\varepsilon}$ with  the properties stated in (\ref{t-p}) it is easy to show
the  asymptotical tightness of the continuity bound (\ref{SBC-ineq}) for large $E$. $\square$ \smallskip

\begin{remark}\label{T1+} By Remark \ref{u-b} all arguments from the proof of Theorem \ref{main} are
valid for any function $f$ satisfying  continuity bounds
(\ref{fcb}) and (\ref{SBC-ineq-gen+}).\medskip
\end{remark}

Assume now that the system $A$  is the $\,\ell$-mode quantum oscillator with the frequencies $\,\omega_1,...,\omega_{\ell}\,$. The Hamiltonian of this system has the form
\begin{equation*}
H_A=\sum_{i=1}^{\ell}\hbar \omega_i a_i^*a_i+E_0 I_A,\quad E_0=\frac{1}{2}\sum_{i=1}^{\ell}\hbar \omega_i,
\end{equation*}
where $a_i$ and $a^*_i$ are the annihilation and creation operators of the $i$-th mode \cite{H-SCI}. Note that this
Hamiltonian satisfies condition (\ref{BD-cond}) with $a=1+1/\ell$ \cite{Datta}.

In this case the function $F_{H_A}(E)$ defined in (\ref{F-def}) is bounded above by the function
\begin{equation}\label{F-ub}
F_{\ell,\omega}(E)\doteq \ell\ln \frac{E+E_0}{\ell E_*}+\ell,\quad E_*=\left[\prod_{i=1}^{\ell}\hbar\omega_i\right]^{1/\ell}\!\!,
\end{equation}
and upper bound (\ref{F-ub}) is $\varepsilon$-sharp for large $E$ \cite{AFM,MCB}. So, the function
\begin{equation}\label{F-ub+}
\bar{F}_{\ell,\omega}(E)\doteq F_{\ell,\omega}(E+E_0)=\ell\ln \frac{E+2E_0}{\ell E_*}+\ell,
\end{equation}
is a  upper bound on the function $\bar{F}_{H_A}(E)\doteq F_{H_A}(E+E_0)$  satisfying all the conditions in (\ref{F-cond-1}),(\ref{F-cond-2}) and (\ref{F-cond-3}) \cite{MCB}.
By using the function $\bar{F}_{\ell,\omega}$ in the role of the function $\hat{F}_{H_A}$
in Theorem \ref{main} we obtain the following\smallskip

\begin{corollary}\label{SCB-G}
\emph{Let $A$ be the $\ell$-mode quantum oscillator with the frequencies $\omega_1,...,\omega_{\ell}$. Let $A^n\doteq A_1...A_n$, where $A_k\cong A$ for $k=\overline{1,m}$, $m\leq n$.
Let $\rho$ and $\sigma$ be arbitrary states in $\S(\H_{A^n})$ such that $\,\sum_{k=1}^{m}\Tr H_{A_k}\rho_{A_k},\,\sum_{k=1}^{m}\Tr H_{A_k}\sigma_{A_k}\leq mE$ and $\;\frac{1}{2}\|\shs\rho-\sigma\|_1\leq\varepsilon$. Let $\,t\in (0, 1/\varepsilon)$. Then
\begin{equation}\label{SBC-ineq+}
\begin{array}{ccc}
\displaystyle|f(\rho)-f(\sigma)|\leq
Cm(\varepsilon+\varepsilon^2t^2)\ell \ln\!\left[\frac{m\bar{E}/(\varepsilon^2t^2)+2E_0}{e^{-1}\ell E_*}\right]
\\\\ \displaystyle+2Cm\sqrt{2\varepsilon t}\shs\ell\ln\!\left[\frac{\bar{E}/(\varepsilon t)+2E_0}{e^{-1}\ell E_*}\right]
+D\!\left(g\!\left(\varepsilon+\varepsilon^2t^2\right)+2g(\sqrt{2\varepsilon t})\right)
\end{array}
\end{equation}
for any function $f$ from the class $\widehat{L}^{m}_n(C,D)$, where $\bar{E}=E-E_0$. Inequality (\ref{SBC-ineq+}) holds for any function $f$ from the class $N^{m}_{n,s}(C,D)$
with $\varepsilon$ replaced by $\sqrt{\varepsilon(2-\varepsilon)}$.}\smallskip

\emph{If $f$ is a function from the class $\widehat{L}^{m}_n(C,D)$ satisfying condition (\ref{at}) then continuity bound (\ref{SBC-ineq+}) with optimal $\,t$ is  asymptotically tight for large $E$ \cite{AT}.}
\end{corollary}

\section{Applications}

\subsection{Multipartite quantum (conditional) mutual information}

The quantum mutual information of a state $\,\rho\,$ of a multipartite system $A_1 \ldots
A_n$ is defined as follows (cf.\cite{L-mi,Herbut,NQD,Y&C,YHW})
\begin{equation}\label{mi-mpd}
     I(A_1\!:\ldots:\!A_n)_{\rho}\doteq
    H(\rho\shs\|\shs\rho_{A_{1}}\otimes\cdots\otimes\rho_{A_{n}})=\sum_{k=1}^n H(\rho_{A_{k}})-H(\rho),
\end{equation}
where the second formula is valid if $H(\rho)<+\infty$. If all the marginal entropies $H(\rho_{A_1}),..., H(\rho_{A_{n}})$ are finite then the second formula in (\ref{mi-mpd}) implies that
\begin{equation}\label{nMI-UB)}
  I(A_1\!:\ldots:\!A_n)_{\rho}\leq \sum_{k=1}^{n}H(\rho_{A_k}).
\end{equation}

It follows from inequality (\ref{F-c-b+}) in Section 2 (with trivial system $C$) that the function $\,f(\rho)=I(A_1\!:\ldots:\!A_n)_{\rho}\,$ satisfies inequality  (\ref{F-p-1}) with $a_f=1$ and $b_f=n-1$. The nonnegativity of the quantum mutual information and upper bound (\ref{nMI-UB)}) show that this function satisfies inequality  (\ref{F-p-2})
with $m=n$, $c^-_f=0$ and $c^+_f=1$. It follows that it belongs to the class $L^{n}_{n}(1,n)$.\footnote{Note that the function $f(\rho)=I(A_1\!:\ldots:\!A_n)_{\rho}$  also satisfies inequality  (\ref{F-p-2})
with $m=n-1$, $c^-_f=0$ and $c^+_f=2$ (see inequality (\ref{nCMI-UB-1)}) with trivial system $C$). So, it also belongs to the class $L^{n-1}_{n}(2,n)$.}

Thus, if all the subsystems $A_1,...,A_n$ are finite-dimensional then Proposition  \ref{f-d-case} implies that
\begin{equation}\label{mi-fcb}
 |I(A_1\!:\ldots:\!A_n)_{\rho}-I(A_1\!:\ldots:\!A_n)_{\sigma}|\leq \varepsilon\ln \dim\H_{A_1...A_n}+ng(\varepsilon)
\end{equation}
for any states $\rho$ and $\sigma$ in $\S(\H_{A_1...A_n})$ such that $\;\frac{1}{2}\|\shs\rho-\sigma\|_1\leq\varepsilon$. If
all the subsystems $A_1,...,A_n$ have the same dimension $d$ then there is a pure state $\rho$ in $\S(\H_{A_1...A_n})$ such that
$\rho_{A_k}=d^{-1}I_{A_k}$ for $k=\overline{1,n}$. Since $I(A_1\!:\ldots:\!A_n)_{\rho}=n\ln d=\ln\dim\H_{A_1...A_n}$, by using any product state
$\sigma$ one can show that continuity bound (\ref{mi-fcb}) is asymptotically tight for large $d$ \cite{AT}. Note that  continuity bound (\ref{mi-fcb}) cannot be obtained
by applying  Audenaert's continuity bound (cf.\cite{Aud}) to the summands in the second formula in (\ref{mi-mpd}).
\smallskip

In the infinite-dimensional case uniform continuity bounds for the function $\,f(\rho)=I(A_1\!:\ldots:\!A_n)_{\rho}\,$ can be obtained by applying Theorems \ref{main-gc} and \ref{main}.
In the following proposition $\,\mathbb{VB}^{m}_{\shs t}(\bar{E},\varepsilon\,|\,C,D)$ denotes  the expression in the r.h.s. of (\ref{SBC-ineq})
defined by means of  any continuous  function $\hat{F}_{H_A}$ on $\mathbb{R}_+$ satisfying conditions (\ref{F-cond-1}) and  (\ref{F-cond-2}).
\smallskip\pagebreak

\begin{property}\label{nMI-CB}  \emph{Let $n\geq2$ be arbitrary and $H_{A_1},...,H_{A_n}$
the Hamiltonians of quantum systems $A_1,...,A_n$ satisfying  condition (\ref{H-cond+}).
Let $\rho$ and $\sigma$ be states in $\,\S(\H_{A_1..A_n})$ such that $\,\sum_{k=1}^{n}\Tr H_{A_k}\rho_{A_k},\,\sum_{k=1}^{n}\Tr H_{A_k}\sigma_{A_k}\leq nE$ and $\,\frac{1}{2}\|\shs\rho-\sigma\|_1\leq\varepsilon\leq 1$. Then
\begin{equation}\label{nMI-CB-1}
|I(A_1\!:\ldots:\!A_n)_{\rho}-I(A_1\!:\ldots:\!A_n)_{\sigma}|\leq \sqrt{2\varepsilon}\shs\bar{F}_{H_{\!A^n}}\!\!\left[\frac{n\bar{E}}{\varepsilon}\right]+ng(\sqrt{2\varepsilon}),
\end{equation}
where $\bar{F}_{H_{\!A^n}}$ is the function defined in (\ref{F-bar}) with $A=A^n\doteq A_1...A_n$ and $\bar{E}=E-E_0^{A^n}\!/n$.}\smallskip

\emph{If $A_k\cong A$ for $k=\overline{1,n}$ then
\begin{equation}\label{nMI-CB-2}
|I(A_1\!:\ldots:\!A_n)_{\rho}-I(A_1\!:\ldots:\!A_n)_{\sigma}|\leq \mathbb{VB}^{n}_{\shs t}(\bar{E},\varepsilon\,|\,1,n)
\end{equation}
for any  $t\in(0,1/\varepsilon)$, where $\bar{E}=E-E^A_0$.}\smallskip

\emph{The right hand sides of (\ref{nMI-CB-1}) and (\ref{nMI-CB-2}) tend to zero as $\shs\varepsilon\to 0$ for given $\bar{E}$ and $\shs t$.}\smallskip

\emph{If conditions (\ref{F-cond-3}) and (\ref{BD-cond}) hold then continuity bound (\ref{nMI-CB-2}) with optimal $\,t$ is asymptotically tight for large $E$ \cite{AT}.
This is true, in particular, if $A$ is the $\ell$-mode quantum oscillator and $\hat{F}_{H_A}=\bar{F}_{\ell,\omega}$.\footnote{The function $\bar{F}_{\ell,\omega}$ is defined in (\ref{F-ub+}).} In this case (\ref{nMI-CB-2}) holds with $\mathbb{VB}^n_{\shs t}(\bar{E},\varepsilon\,|\,1,n)$ replaced by the r.h.s. of (\ref{SBC-ineq+}) with $C=1$ and  $D=n$.}
\end{property}\smallskip

\emph{Proof.} Continuity bounds (\ref{nMI-CB-1}) and (\ref{nMI-CB-2})  follow, respectively, from Theorems \ref{main-gc} and  \ref{main}.
Since the Hamiltonians $H_{\!A_1}$,.., $H_{\!A_{n}}$
satisfy condition (\ref{H-cond+}),  Lemma \ref{sl} implies that $\bar{F}_{H_{\!A^{n}}}(E)$ is $o(\sqrt{E})$ as $E\to+\infty$ and hence the r.h.s. of (\ref{nMI-CB-1})
tends to zero as $\varepsilon\rightarrow0$. The r.h.s. of (\ref{nMI-CB-2}) tends to zero as $\shs\varepsilon\rightarrow0$ by Remark \ref{inc-p}. \smallskip

To prove the asymptotical tightness of continuity bound (\ref{nMI-CB-2}) it suffices, by Theorem \ref{main}, to show that both relations in (\ref{at}) hold for the function $\,f(\rho)=I(A_1\!:\ldots:\!A_n)_{\rho}$.
The first relation in (\ref{at}) can be shown by considering the state $\rho^1_E=\gamma_{A_1}(E)\otimes\cdots\otimes\gamma_{A_n}(E)$ at which the function $f$ is equal to zero for any $E>0$. The second relation in (\ref{at}) can be shown by using the pure state
$$
\rho^2_E=\sum_{i,j}\sqrt{p_ip_j}|\varphi^1_i\rangle\langle\varphi^1_j|\otimes\cdots\otimes|\varphi^n_i\rangle\langle\varphi^n_j|,
$$
where $\,\sum_{i}p_i |\varphi^k_i\rangle\langle\varphi^k_i|\,$ is the spectral decomposition of the Gibbs state $\gamma_{A_k}(E)$ in $\S(\H_{A_k})$,
since it is easy to see that $f(\rho^2_E)=n F_{H_A}(E)$ for any $E>0$.

The last assertion of the proposition follows from Corollary \ref{SCB-G}.
$\square$ \medskip

The quantum conditional mutual information (QCMI) of a state $\,\rho\,$ of a finite-dimensional multipartite system $A_1 \ldots
A_nC$ is defined by conditioning the second expression in (\ref{mi-mpd}), i.e. by replacing all the entropies  $H(\rho_X)$ in this expression by the conditional entropies $H(X|C)_{\rho}$.

Similar to the multipartite  quantum mutual information the multipartite QCMI has a nonnegative lower semicontinuous extension
to the set of all  states of an infinite-dimensional multipartite system $A_1 \ldots
A_nC$ possessing all basic properties of QCMI. But in contrast to the unconditional mutual information the extented multipartite QCMI can not be expressed
by a simple formula for any state in $\S(\H_{A_1..A_nC})$ (see details in Section 2.1).

If the marginal entropies $H(\rho_{A_1}),..., H(\rho_{A_{n-1}})$ of a state $\rho\in\S(\H_{A_1..A_nC})$ are finite then the QCMI is given
by formula (\ref{cmi-mpd+}) in which all the summands are explicitly expressed via the quantum mutual information as follows
$$
I(A_{n-k}\!:\!A_{n-k+1}...A_{n}|C)_{\rho}=I(A_{n-k}\!:\!A_{n-k+1}...A_{n}C)_{\rho}-I(A_{n-k}\!:\!C)_{\rho},\quad k=\overline{1,n-1}.
$$
Upper bound (\ref{CMI-UB}) implies that
\begin{equation}\label{nCMI-UB-1)}
  I(A_1\!:\ldots:\!A_n|C)_{\rho}\leq 2\sum_{k=1}^{n-1}H(\rho_{A_k}).
\end{equation}
If all the marginal entropies $H(\rho_{A_1}),..., H(\rho_{A_{n}})$ are finite then by using
a version of inequality (\ref{nCMI-UB-1)})
with arbitrary $n-1$ subsystems of $A_1...A_n$ (instead of $A_1,...,A_{n-1}$) it is easy to show that
\begin{equation}\label{nCMI-UB-2)}
  I(A_1\!:\ldots:\!A_n|C)_{\rho}\leq 2\,\frac{n-1}{n}\sum_{k=1}^{n}H(\rho_{A_k}).
\end{equation}

It follows from inequality (\ref{F-c-b+}) that the function $\,f(\rho)=I(A_1\!:\ldots:\!A_n|C)_{\rho}\,$ satisfies inequality  (\ref{F-p-1}) with $a_f=1$ and $b_f=n-1$. The nonnegativity of QCMI
and upper bounds (\ref{nCMI-UB-1)}) and (\ref{nCMI-UB-2)}) show that this function  satisfies inequality  (\ref{F-p-2})
with $c^-_f=0$ and $c^+_f=2$ in the case $m=n-1$ and with $c^-_f=0$ and $c^+_f=2-2/n$ in the case $m=n$.
It follows that it belongs to the classes $L^{n-1}_{n}(2,n)$ and $L^{n}_{n}(2-2/n,n)$.\smallskip

Thus, if all the subsystems $A_1,...,A_{n-1}$ are finite-dimensional then Proposition  \ref{f-d-case} implies that
\begin{equation}\label{cmi-fcb}
 |I(A_1\!:\ldots:\!A_n|C)_{\rho}-I(A_1\!:\ldots:\!A_n|C)_{\sigma}|\leq 2\varepsilon\ln \dim\H_{A_1...A_{n-1}}+ng(\varepsilon)
\end{equation}
for any states $\rho$ and $\sigma$ in $\S(\H_{A_1...A_n})$ such that $\;\frac{1}{2}\|\shs\rho-\sigma\|_1\leq\varepsilon$. If all the subsystems $A_1,...,A_{n}$ are finite-dimensional then the first summand in the r.h.s. of (\ref{cmi-fcb}) can be replaced by $(2-2/n)\varepsilon\ln \dim\H_{A_1...A_{n}}$. It is easy to show  that continuity bound (\ref{cmi-fcb}) is asymptotically tight for large $\dim\H_{A_1}$ in the case $n=2$ \cite{AT}.\smallskip

In the infinite-dimensional case we may directly apply Theorems \ref{main-gc} and \ref{main} to the  function $\,f(\rho)=I(A_1\!:\ldots:\!A_n|C)_{\rho}\,$ in both  cases $m=n-1$ and $m=n$. This gives continuity bounds for $I(A_1\!:\ldots:\!A_n|C)_{\rho}$ under two forms of energy constraint:
\begin{itemize}
  \item the energy constraint on the subsystem $A_1...A_{n-1}$;
  \item the energy constraint on the whole system $A_1...A_{n}$.
\end{itemize}
\smallskip

In the following proposition $\,\mathbb{VB}^{m}_{\shs t}(\bar{E},\varepsilon\,|\,C,D)$ denotes  the expression in the r.h.s. of (\ref{SBC-ineq})
defined by means of  any continuous  function $\hat{F}_{H_A}$ on $\mathbb{R}_+$ satisfying conditions (\ref{F-cond-1}) and  (\ref{F-cond-2}).
\smallskip

\begin{property}\label{nCMI-CB}  \emph{Let $n\geq2$ be arbitrary and $H_{A_1},...,H_{A_m}$
the Hamiltonians of quantum systems $A_1,...,A_m$ satisfying  condition (\ref{H-cond+}), where  either $\,m=n-1$ or $\,m=n$.
Let $\rho$ and $\sigma$ be states in $\,\S(\H_{A_1..A_nC})$ such that $\,\sum_{k=1}^{m}\Tr H_{A_k}\rho_{A_k},\,\sum_{k=1}^{m}\Tr H_{A_k}\sigma_{A_k}\leq mE$ and $\,\frac{1}{2}\|\shs\rho-\sigma\|_1\leq\varepsilon\leq1$. Let $C_m=(n-1)/m$ and $A^m\doteq A_1...A_m$. Then
\begin{equation}\label{nCMI-CB-1}
|I(A_1\!:\ldots:\!A_n|C)_{\rho}-I(A_1\!:\ldots:\!A_n|C)_{\sigma}|\leq 2C_m\sqrt{2\varepsilon}\shs\bar{F}_{H_{\!A^m}}\!\!\left[\frac{m\bar{E}}{\varepsilon}\right]+ng(\sqrt{2\varepsilon}),
\end{equation}
where $\bar{F}_{H_{A^m}}$ is the function defined in (\ref{F-bar}) with $A=A^m$ and $\bar{E}=E-E_0^{A^m}\!/m$.}\smallskip

\emph{If $A_k\cong A$ for $k=\overline{1,m}$ then
\begin{equation}\label{nCMI-CB-2}
|I(A_1\!:\ldots:\!A_n|C)_{\rho}-I(A_1\!:\ldots:\!A_n|C)_{\sigma}|\leq \mathbb{VB}^{m}_{\shs t}(\bar{E},\varepsilon\,|\,2C_m,n)
\end{equation}
for any  $t\in(0,1/\varepsilon)$, where $\bar{E}=E-E^A_0$.}\smallskip

\emph{The right hand sides of (\ref{nCMI-CB-1}) and (\ref{nCMI-CB-2}) tends to zero as $\shs\varepsilon\to 0$ for given $\bar{E}$ and $\shs t$.}\smallskip

\emph{If conditions (\ref{F-cond-3}) and (\ref{BD-cond}) hold then continuity bound (\ref{nCMI-CB-2}) with optimal $\,t$ are close-to-tight for large $E$ up to the factor $\,2-2/n\,$ in the main term in both cases $m=n-1$ and $m=n$. This is true, in particular, if $A$ is the $\ell$-mode quantum oscillator and $\hat{F}_{H_A}=\bar{F}_{\ell,\omega}$. In this case (\ref{nCMI-CB-2}) holds with the r.h.s. replaced by the r.h.s. of (\ref{SBC-ineq+}) with $C=2C_m$ and $D=n$.}
\end{property}\smallskip

\emph{Proof.} By the observations before the proposition continuity bounds (\ref{nCMI-CB-1}) and (\ref{nCMI-CB-2})  follow, respectively, from Theorems \ref{main-gc} and  \ref{main}. Since the Hamiltonians $H_{\!A_1}$,.., $H_{\!A_{m}}$
satisfy condition (\ref{H-cond+}),  Lemma \ref{sl} implies that $\bar{F}_{H_{\!A^{m}}}(E)$ is $o(\sqrt{E})$ as $E\to+\infty$ and hence the r.h.s. of (\ref{nCMI-CB-1})
tends to zero as $\varepsilon\rightarrow0$. The r.h.s. of (\ref{nCMI-CB-2}) tends to zero as $\varepsilon\rightarrow0$ by Remark \ref{inc-p}. \smallskip

To prove the assertion concerning accuracy of continuity bound (\ref{nCMI-CB-2}) assume  that conditions (\ref{F-cond-3}) and (\ref{BD-cond}) hold and that $C$ is a trivial system, i.e. $I(A_1\!:\ldots:\!A_n|C)_{\rho}=I(A_1\!:\ldots:\!A_n)_{\rho}$.
By using the states $\rho_E^1$ and $\rho_E^2$ introduced in the proof of Proposition \ref{nMI-CB} and by repeating the arguments from the proof of
the last assertion of Theorem \ref{main} it is easy to show that continuity bound (\ref{nCMI-CB-2}) with optimal $\,t$ is close-to-tight for large $E$ up to the factor $\,2-2/n\,$ in the main term in both cases $m=n-1$ and $m=n$.

The last assertion of the proposition follows from Corollary \ref{SCB-G}.  $\square$\smallskip

\begin{remark}\label{t-r} If $n=2$ and conditions (\ref{F-cond-3}) and (\ref{BD-cond}) hold (in particular, if $A$ is the\break $\ell$-mode quantum oscillator)
then continuity bound (\ref{nCMI-CB-2}) with optimal $\,t$ is asymptotically tight for large $E$ in both cases $m=1$ and $m=2$ \cite{AT}.
\end{remark}\smallskip

Continuity bound (\ref{nCMI-CB-1}) in the case $m=n-1$ implies the following\smallskip

\begin{corollary}\label{CMI-un-cont} \emph{Let $A_1$,...,$A_{n}$ and $C$ be arbitrary quantum systems.}
\emph{If the Hamiltonians $H_{\!A_1}$,.., $H_{\!A_{n-1}}$ satisfy condition (\ref{H-cond+}) then the function $\,\rho\mapsto I(A_1\!:\ldots:\!A_n|C)_{\rho}$ is  uniformly continuous on the set of states $\rho$ in $\S(\H_{A_1...A_nC})$ s.t. $\sum_{k=1}^{n-1}\Tr H_{A_k}\rho_{A_k}\leq E$ for any $E>E_0^{A^{n-1}}$, where $E_0^{A^{n-1}}$ is the sum of minimal eigenvalues of $H_{\!A_1}$,.., $H_{\!A_{n-1}}$.}
\end{corollary}\medskip

\subsection{Squashed entanglement and c-squashed entanglement}

The \emph{squashed entanglement} of a state $\rho$ of a  finite-dimensional multipartite system $A_1...A_n$ is defined as
\begin{equation}\label{mse-def}
  E_{sq}(\rho)=\textstyle\frac{1}{2}\displaystyle\inf_{\hat{\rho}\in \mathfrak{M}_1(\rho)}I(A_1\!:...:\!A_n|E)_{\hat{\rho}},
\end{equation}
where $\mathfrak{M}_1(\rho)$ is the set of all extensions $\hat{\rho}\in\S(\H_{A_1...A_nE})$ of the state
$\rho$ \cite{AHS,C&W,Tucci,Y&C}.\footnote{In \cite{AHS,Y&C} two $n$-partite generalizations of the bipartite squashed entanglement
are proposed: the first one is defined in (\ref{mse-def}), the second one is defined by the expression similar to (\ref{mse-def}) with
the different $n$-partite version of QCMI (called dual conditional total correlation or secrecy monotones). In \cite{D-S-W} it is proved that these $n$-partite
generalizations of the bipartite squashed entanglement coincide.} By using the extended multipartite QCMI described in Section 2.1
this definition can be generalized to any state $\rho$ of an infinite-dimensional $n$-partite system $A_1...A_n$. By using the arguments from \cite{AHS,Y&C} one can show that in this case the function $E_{sq}$
defined by formula (\ref{mse-def}) possesses almost all properties of an entanglement measure, in particular,
it is convex on the whole set of states and nonincreasing under LOCC. Similar to the bipartite case,
it is not clear how to show that $E_{sq}$ is equal to zero on the set of all separable states because of
the existence of countably nondecomposable separable states in infinite-dimensional composite systems (see Remark 10 in \cite{SE}).
\smallskip

The \emph{c-squashed entanglement} of a state $\rho$ of a finite-dimensional multipartite quantum system $A_1...A_n$ is defined as
\begin{equation*}
  E^c_{sq}(\rho)=\textstyle\frac{1}{2}\displaystyle\inf_{\hat{\rho}\in\mathfrak{M}_2(\rho)}I(A_1\!:...:\!A_n|E)_{\hat{\rho}},
\end{equation*}
where $\mathfrak{M}_2(\rho)$ is the set of all extensions $\hat{\rho}\in\S(\H_{A_1...A_nE})$ of the state
$\rho$  having form (\ref{c-ext}) with $A_{n+1}=E$ \cite{N&R,Tucci,Y&C}. By using the extended multipartite QCMI described in Section 2.1
this definition can be generalized to any state $\rho$ of an infinite-dimensional $n$-partite system $A_1...A_n$. The c-squashed entanglement
can be also defined as the mixed convex roof of the quantum mutual information, i.e.
\begin{equation*}
  E^c_{sq}(\rho)=\textstyle\frac{1}{2}\displaystyle\inf_{\sum_i p_i\rho_i=\rho}\sum_i p_i I(A_1\!:...:\!A_n)_{\rho_i},
\end{equation*}
where the infimum is over all countable collections $\{\rho_i\}$ of states in $\S(\H_{A_1...A_n})$ and probability distributions $\{p_i\}$ such that $\sum_i p_i\rho_i=\rho$.
The equality
$$
\sum_i p_i I(A_1\!:...:\!A_n)_{\rho_i}=I(A_1\!:...:\!A_n|E)_{\hat{\rho}},\quad \hat{\rho}=\sum_i p_i \rho_i\otimes |i\rangle\langle i|,
$$
is easily verified if all the states $\rho_i$ have finite marginal entropies and the Shannon entropy of the
probability distribution $\{p_i\}$ is finite. The validity of this equality in general case can be proved by using the approximation property
for the extended multipartite QCMI stated in Proposition 5 in \cite{CMI}.

In Section 4.1 it is mentioned that the function $\rho\mapsto I(A_1\!:...:\!A_n|E)_{\rho}$ belongs to the classes $L^{n-1}_{n}(2,n)$ and $L^{n}_{n}(2-2/n,n)$.
Hence the function $2E_{sq}$ belongs to the classes $N^{n-1}_{n,1}(2,n)$ and $N^{n}_{n,1}(2-2/n,n)$, while
the function $2E^c_{sq}$ belongs to the classes $N^{n-1}_{n,2}(2,n)$ and $N^{n}_{n,2}(2-2/n,n)$. \smallskip

Thus, if the subsystems $A_1,...,A_{n-1}$ are finite-dimensional then Proposition  \ref{f-d-case} implies that
\begin{equation}\label{se-fcb}
 2|E^*_{sq}(\rho)-E^*_{sq}(\sigma)|\leq 2\delta\ln \dim\H_{A_1...A_{n-1}}+ng(\delta),\quad E^*_{sq}=E_{sq},E^c_{sq},
\end{equation}
for any states $\rho$ and $\sigma$ in $\S(\H_{A_1...A_n})$ s.t. $\;\frac{1}{2}\|\shs\rho-\sigma\|_1\leq\varepsilon\leq1$, where $\delta=\sqrt{\varepsilon(2-\varepsilon)}$.
If all the subsystems $A_1,...,A_{n}$ are finite-dimensional then the first summand in the r.h.s. of (\ref{se-fcb}) can be replaced by $(2-2/n)\delta\ln \dim\H_{A_1...A_{n}}$.
\smallskip

In the infinite-dimensional case Theorems \ref{main-gc} and \ref{main} (along with Corollary \ref{SCB-G}) give continuity bounds for the functions  $E_{sq}$ and $E^c_{sq}$ under two forms of energy constraint. They correspond to the cases $m=n-1$ and $m=n$ in the following proposition, in which $\,\mathbb{VB}^{m}_{\shs t}(\bar{E},\varepsilon\,|\,C,D)$ denotes  the expression in the r.h.s. of (\ref{SBC-ineq})
defined by means of  any continuous  function $\hat{F}_{H_A}$ on $\mathbb{R}_+$ satisfying conditions (\ref{F-cond-1}) and  (\ref{F-cond-2}).
\smallskip

\begin{property}\label{nSqE-CB} \emph{Let $\,n\geq2$ be arbitrary and $H_{A_1},...,H_{A_m}$
the Hamiltonians of quantum systems $A_1,..,A_m$ satisfying  condition (\ref{H-cond+}), where  either $\shs m=n-1$ or $\shs m=n$.
Let $\rho$ and $\sigma$ be states in $\,\S(\H_{A_1..A_n})$ such that $\,\sum_{k=1}^{m}\Tr H_{A_k}\rho_{A_k},\,\sum_{k=1}^{m}\Tr H_{A_k}\sigma_{A_k}\leq mE$ and $\,\frac{1}{2}\|\shs\rho-\sigma\|_1\leq\varepsilon\leq 1$. Let $C_m=(n-1)/m$, $A^m\doteq A_1...A_m$ and $\delta=\sqrt{\varepsilon(2-\varepsilon)}$. Then
\begin{equation}\label{nSqE-CB-1}
2|E^*_{sq}(\rho)-E^*_{sq}(\sigma)|\leq 2C_m\sqrt{2\delta}\shs\bar{F}_{H_{\!A^m}}\!\!\left[\frac{m\bar{E}}{\delta}\right]+ng(\sqrt{2\delta}),\quad E^*_{sq}=E_{sq},E^c_{sq},
\end{equation}
where $\bar{F}_{H_{A^m}}$ is the function defined in (\ref{F-bar}) with $A=A^m$ and $\bar{E}=E-E_0^{A^m}/m$.}\smallskip

\emph{If $A_k\cong A$ for $\,k=\overline{1,m}$ then
\begin{equation}\label{nSqE-CB-2}
2|E^*_{sq}(\rho)-E^*_{sq}(\sigma)|\leq \mathbb{VB}^{m}_{\shs t}(\bar{E},\delta\,|\,2C_m,n),\quad E^*_{sq}=E_{sq},E^c_{sq},
\end{equation}
for any  $t\in(0,1/\delta)$, where $\bar{E}=E-E^A_0$.}\smallskip

\emph{The right hand sides of (\ref{nSqE-CB-1}) and (\ref{nSqE-CB-2}) tend to zero as $\shs\varepsilon\to 0$ for given $\bar{E}$ and $\shs t$.}\smallskip

\emph{If $A$ is the $\ell$-mode quantum oscillator then inequality (\ref{nSqE-CB-2}) holds with the r.h.s. replaced by the r.h.s. of (\ref{SBC-ineq+}) with $\delta$ instead of $\varepsilon$, $\,C=2C_m$ and  $D=n$ for any $t\in(0,1/\delta)$.}
\end{property} \medskip

The continuity bounds in (\ref{nSqE-CB-1}) imply the following\smallskip

\begin{corollary}\label{SE-un-cont} \emph{Let $A_1$,...,$A_{n}$ be arbitrary quantum systems.}
\emph{If the Hamiltonians $H_{\!A_1}$,.., $H_{\!A_{n-1}}$ satisfy condition (\ref{H-cond+}) then}\smallskip

\noindent A) \emph{the functions $E_{sq}$ and $E^c_{sq}$ are  uniformly continuous on the set of states $\rho$ in $\S(\H_{A_1...A_n})$ such that  $\,\sum_{k=1}^{n-1}\Tr H_{A_k}\rho_{A_k}\leq E$ for any $E>E_0^{A^{n-1}}\!=E_0^{A_1}+...+E_0^{A_{n-1}}$;}\smallskip

\noindent B) \emph{the functions $E_{sq}$ and $E^c_{sq}$ are asymptotically continuous in the following sense (cf.\cite{ESP}):
if $\{\rho_d\}$ and $\{\sigma_d\}$ are any sequences of states such that
$$
\rho_d,\sigma_d \in\S(\H_{A^d_1...A^d_n}),\quad \Tr H_{\!B^{d}}\rho_{B^{d}}, \Tr H_{\!B^{d}}\sigma_{B^{d}}\leq dE,\;\; \forall d,\quad and\quad  \lim_{d\to+\infty}\|\rho_d-\sigma_d\|_1=0,
$$
where $X^{d}$ denotes $d$ copies of a system $X$, $B=A_1...A_{n-1}$ and $H_{\!B^{d}}$ is the Hamiltonian of the system $B^{d}$, then}
$$
\lim_{d\to+\infty}\frac{|E^*_{sq}(\rho_d)-E^*_{sq}(\sigma_d)|}{d}=0,\quad E^*_{sq}=E_{sq},E^c_{sq}.
$$
\end{corollary}

\emph{Proof.} The first assertion of the corollary  directly follows from the continuity bounds in  (\ref{nSqE-CB-1}) in the case $m=n-1$ (since the r.h.s. of (\ref{nSqE-CB-1}) vanishes as $\varepsilon\to0$). \smallskip

To prove the second assertion note that $F_{H_{\!B^d}}(E)=dF_{H_{\!B}}(E/d)$ and $E_0^{B^d}=dE_0^{B}$  for each $d$ and hence
$\bar{F}_{H_{\!B^d}}(E)=d\bar{F}_{H_{\!B}}(E/d)$. So, continuity bound (\ref{nSqE-CB-1}) with $m=n-1$ implies that
\begin{equation}\label{SE-ucb-k}
    \frac{2|E^*_{sq}(\rho_d)-E^*_{sq}(\sigma_d)|}{d}\leq \displaystyle 2\sqrt{2\delta_d}\bar{F}_{H_{\!B}}\!\!\left(\bar{E}/\delta_d\right)+(n/d)g(\sqrt{2\delta_d}),\quad E^*_{sq}=E_{sq},E^c_{sq},
\end{equation}
where $\delta_d=\sqrt{\varepsilon_d(2-\varepsilon_d)}$, $\varepsilon_d=\frac{1}{2}\|\shs\rho_d-\sigma_d\|_1$ and $\bar{E}=E-E_0^B$. Since the sequence $\{\varepsilon_d\}$ is vanishing by the condition
and $\bar{F}_{H_{\!B}}(E)$ is $o(\sqrt{E})$ as $E\to+\infty$ by Lemma \ref{sl},
the r.h.s. of (\ref{SE-ucb-k}) tends to zero as $d\to+\infty$. $\square$

\subsection{Conditional entanglement of
mutual information}

The \emph{conditional entanglement of
mutual information} of a state $\rho$ of a  finite-dimensional multipartite system $A_1...A_n$ is defined as
\begin{equation*}
  E_{I}(\rho)=\textstyle\frac{1}{2}\displaystyle\inf_{\hat{\rho}\in \mathfrak{M}_1(\rho)}[I(A_1A'_1\!:...:\!A_nA'_n)_{\hat{\rho}}-I(A'_1\!:...:\!A'_n)_{\hat{\rho}}],
\end{equation*}
where $\mathfrak{M}_1(\rho)$ is the set of all extensions $\hat{\rho}\in\S(\H_{A_1..A_nA'_1..A'_n})$ of the state
$\rho$ \cite{CondEnt,YHW}. This definition can be extended to an arbitrary state $\rho$ of an infinite-dimensional multipartite system $A_1...A_n$ by noting that
the function
$$
\Delta(\varrho)=I(A_1A'_1\!:...:\!A_nA'_n)_{\varrho}-I(A'_1\!:...:\!A'_n)_{\varrho}
$$
well defined for any state $\varrho$ with finite $I(A'_1\!:...:\!A'_n)_{\varrho}$  has a nonnegative lower semicontinuous extension
to the set of all states of the infinite-dimensional system $A_1...A_nA'_1...A'_n$ given by the expression
\begin{equation}\label{inf-g-d+}
\Delta(\varrho)=I(A_1\!:\!A'_2...A'_n|A'_1)_{\varrho}+\sum_{k=2}^n
I(A_k\!:\!A_1...A_{k-1}A'_1...A'_{k-1}A'_{k+1}...A'_n|A'_k)_{\varrho},
\end{equation}
in which all the summands are the extended tripartite QCMI described in Section 2.1 \cite[Proposition 8]{CMI}.\footnote{In finite dimensions expression (\ref{inf-g-d+}) was obtained in \cite{NQD}.}
This expression and upper bound (\ref{CMI-UB}) imply that
\begin{equation}\label{D-UB}
\Delta(\varrho)\leq 2\sum_{k=1}^n H(\varrho_{A_k})\quad \forall \varrho\in \S(\H_{A_1...A_nA'_1...A'_n}).
\end{equation}

If $\varrho$ is a state in $\S(\H_{A_1...A_nA'_1...A'_n})$ with finite marginal entropies then
$$
\Delta(\varrho)=\sum_{k=1}^n H(A_k|A'_k)_{\varrho}-H(A_1...A_n|A'_1...A'_n)_{\varrho}.
$$
By using this representation, concavity of the conditional entropy and inequality (\ref{w-k-ineq}) it is easy to show that
the function $f=\Delta$ satisfies inequality (\ref{F-p-1}) with $a_f=1$ and $b_f=n$ for any states $\rho$ and $\sigma$ in
$\S(\H_{A_1...A_nA'_1...A'_n})$ with finite marginal entropies. Using this, representation (\ref{inf-g-d+}) and Corollary 9 in \cite{CMI}
one can prove that the function $f=\Delta$ satisfies inequality (\ref{F-p-1}) with $a_f=1$ and $b_f=n$ for arbitrary states $\rho$ and $\sigma$
in $\S(\H_{A_1...A_nA'_1...A'_n})$. Inequality (\ref{D-UB}) and nonnegativity of $\Delta(\varrho)$ mean that the function $f=\Delta$  satisfies inequality (\ref{F-p-2}) with $c^-_f=0$ and $c^+_f=2$.

These observations show that the function $\Delta$ belongs to the class $L_{2n}^n(2,n+1)$. It follows that the function $2E_I$ belongs
to the class $N^{n}_{n,1}(2,n+1)$.

Thus, if the subsystems $A_1,...,A_{n}$ are finite-dimensional then Proposition  \ref{f-d-case} implies that
\begin{equation*}
 2|E_{I}(\rho)-E_{I}(\sigma)|\leq 2\delta\ln \dim\H_{A_1...A_{n}}+(n+1)g(\delta)
\end{equation*}
for any states $\rho$ and $\sigma$ in $\S(\H_{A_1...A_n})$ s.t. $\;\frac{1}{2}\|\shs\rho-\sigma\|_1\leq\varepsilon\leq1$, where $\delta=\sqrt{\varepsilon(2-\varepsilon)}$. \smallskip

In the infinite-dimensional case continuity bounds for the function $E_{I}$ can be obtained by using Theorems \ref{main-gc} and \ref{main} (along with Corollary \ref{SCB-G}). They are presented in the following proposition, in which $\,\mathbb{VB}^{m}_{\shs t}(\bar{E},\varepsilon\,|\,C,D)$ denotes  the expression in the r.h.s. of (\ref{SBC-ineq})
defined by means of  any continuous  function $\hat{F}_{H_A}$ on $\mathbb{R}_+$ satisfying conditions (\ref{F-cond-1}) and  (\ref{F-cond-2}).
\smallskip

\begin{property}\label{E_I-CB} \emph{Let $\shs n\geq2$ be arbitrary and $H_{A_1},...,H_{A_n}$
the Hamiltonians of quantum systems $A_1,...,A_n$ satisfying  condition (\ref{H-cond+}).
Let $\rho$ and $\sigma$ be states in $\,\S(\H_{A_1..A_n})$ such that $\,\sum_{k=1}^{n}\Tr H_{A_k}\rho_{A_k},\,\sum_{k=1}^{n}\Tr H_{A_k}\sigma_{A_k}\leq nE$ and $\,\frac{1}{2}\|\shs\rho-\sigma\|_1\leq\varepsilon\leq 1$. Let $A^n\doteq A_1...A_n$ and $\delta=\sqrt{\varepsilon(2-\varepsilon)}$. Then
\begin{equation}\label{E_I-CB-1}
2|E_{I}(\rho)-E_{I}(\sigma)|\leq 2\sqrt{2\delta}\shs\bar{F}_{H_{\!A^n}}\!\!\left[\frac{n\bar{E}}{\delta}\right]+(n+1)g(\sqrt{2\delta}),
\end{equation}
where $\bar{F}_{H_{A^n}}$ is the function defined in (\ref{F-bar}) with $A=A^n$ and $\bar{E}=E-E_0^{A^n}/n$.}\smallskip

\emph{If $\,A_k\cong A$ for $\,k=\overline{1,n}\,$ then
\begin{equation}\label{E_I-CB-2}
2|E_{I}(\rho)-E_{I}(\sigma)|\leq \mathbb{VB}^{n}_{\shs t}(\bar{E},\delta\,|\,2,n+1)
\end{equation}
for any  $t\in(0,1/\delta)$, where $\bar{E}=E-E^A_0$.}\smallskip

\emph{The right hand sides of (\ref{E_I-CB-1}) and (\ref{E_I-CB-2}) tend to zero as $\shs\varepsilon\to 0$ for given $\bar{E}$ and $\shs t$.}\smallskip

\emph{If $A$ is the $\ell$-mode quantum oscillator then inequality (\ref{E_I-CB-2}) holds with the r.h.s. replaced by the r.h.s. of (\ref{SBC-ineq+}) with $\delta$ instead of $\varepsilon$, $\,C=2$ and  $D=n+1$ for any $t\in(0,1/\delta)$.}
\end{property} \medskip

Continuity bound (\ref{E_I-CB-1}) imply the following\smallskip

\begin{corollary}\label{I-un-cont} \emph{Let $A_1$,...,$A_{n}$ be arbitrary quantum systems.}
\emph{If the Hamiltonians $H_{\!A_1}$,.., $H_{\!A_{n}}$ satisfy condition (\ref{H-cond+}) then}\smallskip

\noindent A) \emph{the function $E_{I}$ is  uniformly continuous on the set of states $\rho$ in $\S(\H_{A_1...A_n})$ such that  $\,\sum_{k=1}^{n}\Tr H_{A_k}\rho_{A_k}\leq E$ for any $E>E_0^{A^{n}}\!=E_0^{A_1}+...+E_0^{A_{n}}$;}\smallskip

\noindent B) \emph{the function $E_{I}$ is asymptotically continuous in the following sense (cf.\cite{ESP}):
if $\{\rho_d\}$ and $\{\sigma_d\}$ are any sequences of states such that
$$
\rho_d,\sigma_d \in\S(\H_{A^d_1...A^d_n}),\quad \Tr H_{\!B^{d}}\rho_d, \Tr H_{\!B^{d}}\sigma_d\leq dE,\;\; \forall d,\quad and\quad  \lim_{d\to+\infty}\|\rho_d-\sigma_d\|_1=0,
$$
where $X^{d}$ denotes $d$ copies of a system $X$, $B=A_1...A_{n}$ and $H_{\!B^{d}}$ is the Hamiltonian of the system $B^{d}$, then}
$$
\lim_{d\to+\infty}\frac{|E_{I}(\rho_d)-E_{I}(\sigma_d)|}{d}=0.
$$
\end{corollary}\medskip

\emph{Proof.} The first assertion of the corollary  directly follows from continuity bound (\ref{E_I-CB-1}), since the r.h.s. of  (\ref{E_I-CB-1}) vanishes as $\varepsilon\to0$. \smallskip

The second  assertion is derived from continuity bound (\ref{E_I-CB-1}) by repeating the arguments from the proof of Corollary \ref{SE-un-cont}. $\square$

\section{On preserving continuity bounds under local channels}

Many characteristics of a multipartite quantum system $A_1...A_n$ are nonnegative and do not increase under
actions of local channels, i.e. channels of the form
\begin{equation}\label{l-ch}
\Lambda=\Phi_1\otimes\Phi_2\otimes\cdots\otimes\Phi_n,
\end{equation}
where $\Phi_k$ is a channel from the system $A_k$ to any system $A_k'$, $k=\overline{1,n}$.

Assume now that $f$ is any function on
$\S(\H_{A_1...A_n})$ possessing the above properties and satisfying inequalities  (\ref{F-p-1}) and (\ref{F-p-2}).
Since a quantum channel is a linear map, it follows that for any local channel $\Lambda:A_1...A_n\rightarrow A'_1...A'_n$ the function $f\circ\Lambda$
also satisfyes inequalities  (\ref{F-p-1}) and (\ref{F-p-2}) with the same parameters.\footnote{We assume here that the function $f$ is defined on the set of states of any $n$-partite system, in particular, the system $A'_1...A'_n$} In terms of the classes introduced in Section 3 this means that
$$
f\in\widehat{L}_{n}^m(C,D) \qquad \Rightarrow \qquad f\circ\Lambda\in\widehat{L}_{n}^m(C,D).
$$
So, by applying Proposition \ref{f-d-case}, Theorem \ref{main-gc} and Theorem \ref{main} to the function
$f\circ\Lambda$ we obtain the same continuity bound for $f\circ\Lambda$ as for the function $f$. \smallskip

For example, the nonnegativity and monotonicity of the quantum mutual information under local channels implies the following
\smallskip

\begin{property}\label{cbp-1} \emph{Let $\,\Lambda:\T(\H_{A_1..A_n})\rightarrow\T(\H_{A'_1..A'_n})$ be a channel having form (\ref{l-ch})}.

A) \emph{If all the subsystems $A_1,...,A_n$ are finite-dimensional then
\begin{equation}\label{mi-fcb+}
|I(A'_1\!:\ldots:\!A'_n)_{\Lambda(\rho)}-I(A'_1\!:\ldots:\!A'_n)_{\Lambda(\sigma)}|\leq \varepsilon\ln \dim\H_{A_1...A_n}+ng(\varepsilon)
\end{equation}
for any states $\rho$ and $\sigma$ in $\S(\H_{A_1...A_n})$ such that $\;\frac{1}{2}\|\shs\rho-\sigma\|_1\leq\varepsilon$.}\smallskip

B) \emph{If the assumptions of Proposition \ref{nMI-CB} hold then inequalities (\ref{nMI-CB-1}) and (\ref{nMI-CB-2})
remain valid with the left hand side replaced by}
$$
|I(A'_1\!:\ldots:\!A'_n)_{\Lambda(\rho)}-I(A'_1\!:\ldots:\!A'_n)_{\Lambda(\sigma)}|.
$$
\end{property}\smallskip

Note that inequality (\ref{mi-fcb+}) holds regardless of the dimensions of the subsystems $A'_1,...,A'_n$ (which may be infinite).
\smallskip

Note also that the assertion of Proposition \ref{cbp-1} remains valid  for any positive trace preserving linear map $\Lambda:\T(\H_{A_1..A_n})\rightarrow\T(\H_{A'_1..A'_n})$ such that
$$
I(A'_1\!:\ldots:\!A'_n)_{\Lambda(\rho)}\leq I(A_1\!:\ldots:\!A_n)_{\rho}\quad \textrm{for any}\;\; \rho\in \S(\H_{A_1..A_n}).
$$

Proposition \ref{cbp-1} states, roughly speaking, that the continuity bound for the quantum mutual information given by Proposition \ref{nMI-CB} is preserved
by local channels. Similar assertion holds for both continuity bounds for the QCMI given by Proposition \ref{nCMI-CB}. \medskip

\textbf{Concluding remarks.} We have proposed  universal methods for quantitative continuity analysis of characteristics of
multipartite quantum systems. The limited size of the article allowed us to consider only several  applications
of these methods. In fact, they can be applied to many other characteristics of multipartite quantum systems, including
the relative entropy of entanglement, conditional and unconditional dual total correlation  \cite{Han78} (also called secrecy monotones \cite{CMS02,Y&C}), the interaction information of a $n$-partite quantum system  (the topological entanglement entropy in the case $n=3$) \cite{J&B,K&P}, etc.
\bigskip

I am grateful to A.S.Holevo and G.G.Amosov for the discussion that motivated this research.
I am also grateful to S.N.Filippov and K.Zyczkowski for useful references.

\end{document}